\documentclass[thmsa,11pt]{article}
\usepackage{amssymb}
\usepackage{amsmath}
\usepackage{amsfonts}

\numberwithin{equation}{section}

\begin{document}

\title{\textbf{Conserved charges for black holes in Einstein-Gauss-Bonnet
gravity coupled to nonlinear electrodynamics in AdS space}}
\author{Olivera Mi\v{s}kovi\'{c}$\,^{a,b}$ and Rodrigo Olea$\,^{a}\,$%
\medskip \\
%EndAName
$^{a}${\small \emph{Instituto de F\'{\i}sica, Pontificia Universidad Cat\'{o}%
lica de Valpara\'{\i}so, Casilla 4059, Valpara\'{\i}so, Chile.}}\\
$^{b}$\emph{{\small Max-Planck-Institut f\"{u}r Gravitationsphysik,
Albert-Einstein-Institut,}}\\
\emph{{\small Am M\"{u}hlenberg 1, 14476 Golm, Germany.}}{\small \emph{%
\medskip }}\\
{\small olivera.miskovic@ucv.cl, \ rodrigo.olea@ucv.cl}}
\maketitle

\begin{abstract}
Motivated by possible applications within the framework of anti-de Sitter
gravity/Conformal Field Theory (AdS/CFT) correspondence, charged black holes
with AdS asymptotics, which are solutions to Einstein-Gauss-Bonnet gravity
in $D$ dimensions, and whose electric field is described by a nonlinear
electrodynamics (NED) are studied.

For a topological static black hole ansatz, the field equations are exactly
solved in terms of the electromagnetic stress tensor for an arbitrary NED
Lagrangian, in any dimension $D$ and for arbitrary positive values of
Gauss-Bonnet coupling. In particular, this procedure reproduces the black
hole metric in Born-Infeld and conformally invariant electrodynamics
previously found in the literature. Altogether, it extends to $D>4$ the
four-dimensional solution obtained by Soleng in logarithmic electrodynamics,
which comes from vacuum polarization effects.

Fall-off conditions for the electromagnetic field that ensure the finiteness
of the electric charge are also discussed. The black hole mass and vacuum
energy as conserved quantities associated to an asymptotic timelike Killing
vector are computed using a background-independent regularization of the
gravitational action based on the addition of counterterms which are a given
polynomial in the intrinsic and extrinsic curvatures.
\end{abstract}

\section{Introduction}

Gauge theories which are described by a nonlinear action for Abelian or
non-Abelian fields have become standard in the context of superstring
theory. Indeed, it was proposed in Ref.\cite{Fradkin-Tseytlin} that all
order loop corrections to gravity should be summed up as a Born-Infeld (%
\textbf{BI}) type Lagrangian \cite{Born-Infeld}. Furthermore, the dynamics
of D-branes is given in terms of a non-Abelian Born-Infeld action \cite%
{Leigh}.

On the other hand, coupling nonlinear electrodynamics (\textbf{NED}) to
gravity has been considered in the literature as a plausible mechanism to
obtain regular black hole solutions (see, for instance, \cite%
{Ayon-Beato}). In this respect, the metric for static,
spherically symmetric black holes for the BI theory minimally coupled to
Einstein gravity was derived in a number of papers \cite%
{Hoffman-Gibbons-Rasheed,Oliveira}. Other gravitating NED models supporting
electrically charged black hole solutions have been also investigated, e.g.,
in Ref.\cite{Heisenberg-Euler} for the Euler-Heisenberg effective Lagrangian
of QED, in Ref.\cite{Soleng} for a logarithmic Lagrangian, and in Ref.\cite%
{Maeda-Hassaine-Martinez} for a Lagrangian defined as powers of the Maxwell
term. In the same spirit, as an example of lower-dimensional models, it is
worth mentioning the study of black holes generated by Coulomb-like fields
in $(2+1)$ dimensions \cite{Cataldo-Cruz-Campo-Garcia}, and a similar
treatment which includes torsion in Ref.\cite{Blagojevic-Cvetkovic-Miskovic}.

Within the framework of AdS/CFT correspondence, higher-derivative
corrections to either gravitational or electromagnetic action in AdS space
are expected to modify the dynamics of the strongly coupled dual theory. In
particular, in hydrodynamic models, the addition of $R^{2}$ terms changes
the ratio of shear viscosity over entropy density \cite{viscosity},
violating the universal bound $1/4\pi $ proposed in Ref.\cite{KSS}. In turn,
it has been proved that higher-derivative terms for Abelian fields in the
form of NED do not affect this ratio \cite{Cai-Sun} (for hydrodynamic models dual
to R-charged black holes see, e.g., Ref.\cite{Ge et al}). Also, in applications
of the AdS/CFT conjecture to high $T_{c}$ superconductivity, higher
curvature terms violate a universal relation between the critical
temperature of the superconductor and its energy gap \cite%
{Gregory-Kanno-Soda,Pan-Wang-Papantonopoulos-Oliveira-Pavan}. While the
Gauss-Bonnet term makes the condensation easier, the inclusion of
Born-Infeld electrodynamics produces the opposite effect \cite{Jing-Chen}.

Motivated by the recent results mentioned above, we study black hole
solutions in Einstein-Gauss-Bonnet gravity with negative cosmological
constant coupled to an arbitrary NED theory. As it is required in the
context of AdS/CFT, we provide definitions for the conserved quantities
following a background-independent regularization procedure.

\section{Action and equations of motion}

We consider a fully-interacting theory of gravity minimally coupled to
nonlinear electrodynamics in a $D$-dimensional manifold $\mathcal{M}$, which
comes from the action
\begin{equation}
I_{0}=\int\limits_{\mathcal{M}}d^{D}x\,\sqrt{-g}\,\mathcal{L}%
_{0}=I_{grav}+I_{NED}\,.  \label{bulk action}
\end{equation}%
The pure gravity part of the bulk action with the metric $g_{\mu \nu }(x)$
as the dynamic field is given by%
\begin{equation}
I_{grav}=\frac{1}{16\pi G}\int\limits_{\mathcal{M}}d^{D}x\,\sqrt{-g}\,\left[
R-2\Lambda +\alpha \,\left( R^{2}-4R_{\mu \nu }R^{\mu \nu }+R_{\mu \nu
\lambda \sigma }R^{\mu \nu \lambda \sigma }\right) \right] \,,
\end{equation}%
which contains the Einstein-Hilbert (\textbf{EH}) action --linear in the
curvature of spacetime--, a cosmological term and a quadratic curvature
correction given by the Gauss-Bonnet (\textbf{GB}) term. The cosmological
constant $\Lambda $ is expressed in terms of the AdS radius $\ell $ as $%
\Lambda =-\left( D-1\right) \left( D-2\right) /2\ell ^{2}$ and $G$ is the
gravitational constant. The GB coupling constant $\alpha $ is of dimension
[length]$^{2}$, which takes only positive values and it is related to the
Regge slope parameter or string scale.

The matter and its interaction with gravity are described by an
electrodynamics action which is nonlinear in the quadratic term $%
F^{2}=g^{\mu \lambda }g^{\nu \rho }F_{\mu \nu }F_{\lambda \rho }$, where $%
F_{\mu \nu }(x)$ is the Abelian field strength associated to the gauge
connection $A_{\mu }(x)$ as $F_{\mu \nu }=\partial _{\mu }A_{\nu }-\partial
_{\nu }A_{\mu }$. We shall assume an action for nonlinear electrodynamics of
the form
\begin{equation}
I_{NED}=\int\limits_{\mathcal{M}}d^{D}x\,\sqrt{-g}\,\mathcal{L}(F^{2})\,,
\label{NED}
\end{equation}%
where the Lagrangian density $\mathcal{L}(F^{2})$ is an arbitrary function
of $F^{2}$.

We will consider the spacetimes whose dimension is $D>4$. The case $D=4$\ is
special because the Euler-Gauss-Bonnet term becomes a topological invariant
that does not contribute to the equations of motion. In that sense, bulk
dynamics in $D=4$ leaves the GB coupling as completely arbitrary. It is
expected, however, that the GB term would modify the boundary dynamics of
the theory and the value of the Euclidean continuation of the action.
Indeed, in four-dimensional AdS gravity, the only consistent way of
achieving the finiteness of both the conserved current and the Euclidean
action is setting $\alpha =\ell ^{2}/4$. Furthermore, nonlinear
electrodynamics in four dimensions is somewhat particular, because one can
consider a Lagrangian that depends additionally on another quadratic
invariant\ $F^{\ast }F=\frac{1}{\sqrt{-g}}\,\epsilon ^{\mu \nu \lambda
\sigma }F_{\mu \nu }F_{\lambda \sigma }$, which by itself is a topological
term. For a recent discussion on electrostatic configurations in
four-dimensional gravitating NED, see Ref.\cite{Diaz-Rubiera}. This type of
Lagrangians clearly cannot be generalized to the\textbf{\ }%
higher-dimensional cases we are interested in.

In order to find the equations of motion of Einstein-Gauss-Bonnet (\textbf{%
EGB}) gravity, we first note that the gravitational action can be rearranged
as%
\begin{eqnarray}
I_{grav} &=&\frac{1}{16\pi G\left( D-2\right) \left( D-3\right) }%
\int\limits_{\mathcal{M}}d^{D}x\,\sqrt{-g}\,\delta _{\lbrack \nu _{1}\cdots
\nu _{4}]}^{[\mu _{1}\cdots \mu _{4}]}\,\left( \frac{1}{2}\,R_{\mu _{1}\mu
_{2}}^{\nu _{1}\nu _{2}}\,\delta _{\mu _{3}}^{\nu _{3}}\delta _{\mu
_{4}}^{\nu _{4}}\right.  \notag \\
&&\qquad \qquad \qquad +\left. \frac{D-2}{D\,\ell ^{2}}\,\delta _{\mu
_{1}}^{\nu _{1}}\delta _{\mu _{2}}^{\nu _{2}}\delta _{\mu _{3}}^{\nu
_{3}}\delta _{\mu _{4}}^{\nu _{4}}+\frac{\alpha \left( D-2\right) \left(
D-3\right) }{4}\,R_{\mu _{1}\mu _{2}}^{\nu _{1}\nu _{2}}R_{\mu _{3}\mu
_{4}}^{\nu _{3}\nu _{4}}\right) \,,  \label{EH-GB}
\end{eqnarray}%
where the tensor $\delta _{\lbrack \nu _{1}\cdots \nu _{p}]}^{[\mu
_{1}\cdots \mu _{p}]}$ denotes the totally antisymmetric product of $p$
Kronecker deltas (see Appendix \ref{Delta}) and we have used the identity%
\begin{equation}
R^{2}-4R_{\mu \nu }R^{\mu \nu }+R_{\mu \nu \lambda \sigma }R^{\mu \nu
\lambda \sigma }=\frac{1}{4}\,\delta _{\lbrack \nu _{1}\cdots \nu
_{4}]}^{[\mu _{1}\cdots \mu _{4}]}\,R_{\mu _{1}\mu _{2}}^{\nu _{1}\nu
_{2}}R_{\mu _{3}\mu _{4}}^{\nu _{3}\nu _{4}}\,.
\end{equation}%
This is a convenient form to take the variation of the Riemann tensor as%
\begin{equation*}
\delta R_{\;\ \nu \alpha \beta }^{\mu }=\nabla _{\alpha }(\delta \Gamma
_{\nu \beta }^{\mu })-\nabla _{\beta }(\delta \Gamma _{\nu \alpha }^{\mu })
\end{equation*}%
in terms of the Christoffel symbol. In addition, using the Bianchi identity
for the Riemann curvature,
\begin{equation*}
\nabla _{\lbrack \mu }R_{\nu \lambda ]}^{\alpha \beta }=\nabla _{\mu }R_{\nu
\lambda }^{\alpha \beta }+\nabla _{\lambda }R_{\mu \nu }^{\alpha \beta
}+\nabla _{\nu }R_{\lambda \mu }^{\alpha \beta }=0\,,
\end{equation*}%
one can show that the gravitational action changes under an arbitrary
variation of the metric as%
\begin{equation}
\delta I_{grav}=-\frac{1}{16\pi G}\int\limits_{\mathcal{M}}d^{D}x\,\sqrt{-g}%
\,\left( g^{-1}\delta g\right) _{\mu }^{\nu }\left( G_{\nu }^{\mu }+H_{\nu
}^{\mu }\right) +\int\limits_{\partial \mathcal{M}}d^{D-1}x\,\Theta
_{grav}(\delta g,\delta \Gamma )\,,  \label{varIgrav}
\end{equation}%
where $G_{\nu }^{\mu }$ is the Einstein tensor with cosmological constant%
\begin{equation}
G_{\nu }^{\mu }=R_{\nu }^{\mu }-\frac{1}{2}\,\delta _{\nu }^{\mu }R+\Lambda
\,\delta _{\nu }^{\mu }\,,
\end{equation}%
and the contribution of the GB term to the variation of the bulk action is
expressed in terms of the Lanczos tensor%
\begin{eqnarray}
H_{\nu }^{\mu } &=&-\frac{\alpha }{8}\,\delta _{\lbrack \nu \nu _{1}\cdots
\nu _{4}]}^{[\mu \mu _{1}\cdots \mu _{4}]}\,R_{\mu _{1}\mu _{2}}^{\nu
_{1}\nu _{2}}R_{\mu _{3}\mu _{4}}^{\nu _{3}\nu _{4}}\,,  \label{H} \\
&=&-\frac{\alpha }{2}\,\delta _{\nu }^{\mu }\left( R^{2}-4R^{\alpha \beta
}R_{\alpha \beta }+R^{\alpha \beta \lambda \sigma }R_{\alpha \beta \lambda
\sigma }\right)  \notag \\
&&+2\alpha \left( RR_{\nu }^{\mu }-2R^{\mu \lambda }R_{\lambda \nu
}-2R_{\lambda \nu \sigma }^{\mu }R^{\lambda \sigma }+R^{\mu \alpha \lambda
\sigma }R_{\nu \alpha \lambda \sigma }\right) \,.
\end{eqnarray}%
The boundary term in (\ref{varIgrav}) that appears from the variation of the
bulk action reads
\begin{equation}
\int\limits_{\partial \mathcal{M}}d^{D-1}x\,\Theta _{grav}=-\frac{1}{16\pi G%
}\int\limits_{\mathcal{M}}d^{D}x\,\partial _{\mu }\left[ \sqrt{-g}\,\delta
_{\lbrack \nu \nu _{1}\nu _{2}\nu _{3}]}^{[\mu \mu _{1}\mu _{2}\mu
_{3}]}\,g^{\nu \alpha }\delta \Gamma _{\mu _{1}\alpha }^{\nu _{1}}\left(
\alpha R_{\mu _{2}\mu _{3}}^{\nu _{2}\nu _{3}}+\frac{1}{\left( D-2\right)
\left( D-3\right) }\,\delta _{\mu _{2}}^{\nu _{2}}\delta _{\mu _{3}}^{\nu
_{3}}\right) \right] .  \label{BT_EGB}
\end{equation}

On the other hand, arbitrary variations of the metric and the gauge field $%
A_{\mu }$ in the NED action produce
\begin{equation}
\delta I_{NED}=\int\limits_{\mathcal{M}}d^{D}x\sqrt{-g}\left[ \frac{1}{2}%
\,T_{\nu }^{\mu }\,\left( g^{-1}\delta g\right) _{\mu }^{\nu }-4\nabla _{\mu
}\left( \frac{d\mathcal{L}}{dF^{2}}\,F^{\mu \nu }\right) \delta A_{\nu }%
\right] +\int\limits_{\partial \mathcal{M}}d^{D-1}x\,\Theta _{NED}(\delta
A)\,,
\end{equation}%
upon a suitable use of the Bianchi identity for the field strength, $%
\partial _{\lbrack \mu }F_{\nu \lambda ]}=\partial _{\mu }F_{\nu \lambda
}+\partial _{\mu }F_{\nu \lambda }+\partial _{\mu }F_{\nu \lambda }=0$. The
energy-momentum tensor for the matter content, $T^{\mu \nu }=\frac{2}{\sqrt{%
-g}}\frac{\delta I_{NED}}{\delta g_{\mu \nu }}$, has the form%
\begin{equation}
T_{\nu }^{\mu }=\delta _{\nu }^{\mu }\,\mathcal{L}-4\,\frac{d\mathcal{L}}{%
dF^{2}}\,F^{\mu \lambda }F_{\nu \lambda }\,,
\end{equation}%
and the surface term of the electromagnetic part is%
\begin{equation}
\int\limits_{\partial \mathcal{M}}d^{D-1}x\,\Theta _{NED}=4\int\limits_{%
\mathcal{M}}d^{D}x\,\partial _{\mu }\left( \sqrt{-g}\,\frac{d\mathcal{L}}{%
dF^{2}}\,F^{\mu \nu }\delta A_{\nu }\right) \,.  \label{BT_matter}
\end{equation}

The variation of the total action (\ref{bulk action}) leads to the field
equations plus a surface term%
\begin{equation}
\delta I_{0}=-\int\limits_{\mathcal{M}}d^{D}x\,\sqrt{-g}\left[ \frac{1}{%
16\pi G}\,\mathcal{E}_{\nu }^{\mu }\,\left( g^{-1}\delta g\right) _{\mu
}^{\nu }+4\,\mathcal{E}^{\mu }\,\delta A_{\mu }\right] +\int\limits_{%
\partial \mathcal{M}}d^{D-1}x\,\Theta _{0}(\delta g,\delta \Gamma ,\delta A)%
\text{\thinspace },  \label{varI}
\end{equation}%
where $\Theta _{0}$\ is the total boundary term coming from the variation of
the bulk action, i.e., $\Theta _{0}=\Theta _{grav}+\Theta _{NED}$.

The equations of motion are then obtained as $\delta I_{0}/\delta g_{\mu \nu
}=0$ and $\delta I_{0}/\delta A_{\mu }=0$, that is,%
\begin{eqnarray}
\mathcal{E}_{\nu }^{\mu } &\equiv &G_{\nu }^{\mu }+H_{\nu }^{\mu }-8\pi
G\,T_{\nu }^{\mu }=0\,,  \label{E^mu_nu} \\
\mathcal{E}^{\mu } &\equiv &\nabla _{\nu }\left( F^{\mu \nu }\frac{d\mathcal{%
L}}{dF^{2}}\right) =0\,.  \label{E_mu}
\end{eqnarray}%
In general, the extremization of the action for the fully-interacting theory
does not only require the e.o.m to be satisfied, but also the vanishing of
the surface term for given boundary conditions. Therefore, a well-posed
action principle leads to supplementing the Lagrangian by suitable boundary
terms, what\textbf{\ }will be discussed below.

The Einstein tensor $G_{\nu }^{\mu }$ can be conveniently rewritten in terms
of the AdS radius as%
\begin{equation}
G_{\nu }^{\mu }=-\frac{1}{4}\,\delta _{\lbrack \nu \nu _{1}\nu _{2}]}^{[\mu
\mu _{1}\mu _{2}]}\left( R_{\mu _{1}\mu _{2}}^{\nu _{1}\nu _{2}}+\frac{1}{%
\ell ^{2}}\,\delta _{\lbrack \mu _{1}\mu _{2}]}^{[\nu _{1}\nu _{2}]}\right)
\,.
\end{equation}%
Written in this compact form, the total equation of motion (\ref{E^mu_nu}) is%
\begin{equation}
\mathcal{E}_{\nu }^{\mu }=-\frac{1}{8}\,\delta _{\lbrack \nu \nu _{1}\cdots
\nu _{4]}}^{[\mu \mu _{1}\cdots \mu _{4}]}\,\left[ \alpha R_{\mu _{1}\mu
_{2}}^{\nu _{1}\nu _{2}}R_{\mu _{3}\mu _{4}}^{\nu _{3}\nu _{4}}+\frac{1}{%
\left( D-3\right) \left( D-4\right) }\,\left( R_{\mu _{1}\mu _{2}}^{\nu
_{1}\nu _{2}}\,\delta _{\lbrack \mu _{3}\mu _{4}]}^{[\nu _{3}\nu _{4}]}+%
\frac{1}{\ell ^{2}}\,\delta _{\lbrack \mu _{1}\mu _{2}]}^{[\nu _{1}\nu
_{2}]}\,\delta _{\lbrack \mu _{3}\mu _{4}]}^{[\nu _{3}\nu _{4}]}\right) %
\right] -8\pi G\,T_{\nu }^{\mu }\,.
\end{equation}%
The GB contribution $H_{\nu }^{\mu }$ given by (\ref{H}) modifies the
cosmological constant in $G_{\nu }^{\mu }$ and therefore, the asymptotic
behavior of the solutions. This is particularly evident in absence of matter
fields, by taking the condition of maximally symmetric spacetimes with an
effective AdS radius $\ell _{e\!f\!f}$, i.e.,
\begin{equation}
R_{\mu \nu }^{\alpha \beta }=-\frac{1}{\ell _{e\!f\!f}^{2}}\,\delta _{\left[
\mu \nu \right] }^{\left[ \alpha \beta \right] }.  \label{LAdSleff}
\end{equation}%
The vacua of the theory are then solutions of global constant curvature,
where $\ell _{e\!f\!f}^{2}$ is a root of the quadratic equation%
\begin{equation}
\alpha \left( D-3\right) \left( D-4\right) \,\frac{1}{\ell _{e\!f\!f}^{4}}-%
\frac{1}{\ell _{e\!f\!f}^{2}}+\frac{1}{\ell ^{2}}=0\,,
\end{equation}%
so that%
\begin{equation}
\ell _{e\!f\!f}^{(\pm )2}=\frac{2\alpha \left( D-3\right) \left( D-4\right)
}{1\pm \sqrt{1-\frac{4\alpha }{\ell ^{2}}\,\left( D-3\right) \left(
D-4\right) }}\,,\qquad \alpha \leq \frac{\ell ^{2}}{4\left( D-3\right)
\left( D-4\right) }\,.  \label{l_eff pm}
\end{equation}%
The GB term, therefore, sets the equations of motion in the
quadratic-curvature form
\begin{equation}
-\frac{\alpha }{8}\,\delta _{\lbrack \nu \nu _{1}\cdots \nu _{4]}}^{[\mu \mu
_{1}\cdots \mu _{4}]}\,\left( R_{\mu _{1}\mu _{2}}^{\nu _{1}\nu _{2}}+\frac{1%
}{\ell _{e\!f\!f}^{(+)2}}\,\delta _{\lbrack \mu _{1}\mu _{2}]}^{[\nu _{1}\nu
_{2}]}\right) \left( R_{\mu _{3}\mu _{4}}^{\nu _{3}\nu _{4}}+\frac{1}{\ell
_{e\!f\!f}^{(-)2}}\,\delta _{\lbrack \mu _{3}\mu _{4}]}^{[\nu _{3}\nu
_{4}]}\right) =8\pi G\,T_{\nu }^{\mu }\,.
\end{equation}%
For the discussion of the present paper, we shall consider solutions that
satisfy the condition (\ref{LAdSleff}) in the asymptotic region, i.e., tend
asymptotically to a constant-curvature spacetime.

However, for different roots $\ell _{e\!f\!f}^{(+)2}\neq \ell
_{e\!f\!f}^{(-)2}$, there is only one branch of the theory of physical
interest. This is because the corresponding AdS radii can be expanded as%
\begin{eqnarray}
\ell _{e\!f\!f}^{(+)2} &=&\alpha \left( D-3\right) \left( D-4\right) +%
\mathcal{O}(\alpha ^{2})\,, \\
\ell _{e\!f\!f}^{(-)2} &=&\ell ^{2}+\mathcal{O}(\alpha )\,,
\end{eqnarray}%
and, thus, $\ell _{e\!f\!f}^{(-)2}$ reduces to the original AdS radius for
vanishing GB coupling, whereas\ $\ell _{e\!f\!f}^{(+)2}$ vanishes if the GB
term goes to zero.

EGB AdS gravity possesses a unique AdS vacuum when both effective AdS radii
are equal, $\ell _{e\!f\!f}^{(+)2}=\ell _{e\!f\!f}^{(-)2}=\ell ^{2}/2$, case
that corresponds to a GB coupling given by $\alpha =\ell ^{2}/4\left(
D-3\right) \left( D-4\right) $. In five dimensions, at that particular
coupling value, the action features a group symmetry enhancement from local
Lorentz to AdS$_{5}$, and it can be expressed as a Chern-Simons density for
the latter group. This gravity theory has particular dynamical features that
will not be discussed here \cite%
{Banados-Olea-Theisen,Banados-Miskovic-Theisen}.

\section{Generic topological static black hole\textbf{\ solution}}

A static black hole ansatz for the metric $g_{\mu \nu }$ in the coordinate
set $x^{\mu }=(t,r,\varphi ^{m})$ is given by%
\begin{equation}
ds^{2}=g_{\mu \nu }(x)\,dx^{\mu }dx^{\nu }=-f^{2}(r)\,dt^{2}+\frac{dr^{2}}{%
f^{2}(r)}+r^{2}\gamma _{mn}(\varphi )\,d\varphi ^{m}d\varphi ^{n}\,.
\label{BH}
\end{equation}%
The boundary $\partial \mathcal{M}$ is located at radial infinity ($%
r\rightarrow \infty $), and it is parameterized by $x^{i}=(t,\varphi ^{m})$.
The metric $\gamma _{nm}$ with local coordinates $\varphi ^{m}$ describes a $%
(D-2)$-dimensional Riemann space $\Gamma _{D-2}$ with constant curvature,
that is,
\begin{equation}
\mathcal{\tilde{R}}_{m_{1}m_{2}n_{1}n_{2}}(\gamma )=k\left( \gamma
_{m_{1}n_{1}}\gamma _{m_{2}n_{2}}-\gamma _{m_{1}n_{2}}\gamma
_{m_{2}n_{1}}\right) \,,
\end{equation}%
where $k=0,$ $+1$ or $-1$, that corresponds to flat, spherical or hyperbolic
transversal section, respectively.

We will consider that the solution possesses an event horizon, defined as
the largest root of the equation $f(r_{+})=0$. The non-vanishing components
of the Riemann curvature $R_{\lambda \rho }^{\mu \nu }$ are%
\begin{eqnarray}
R_{tr}^{tr} &=&-\dfrac{1}{2}\,\left( f^{2}\right) ^{\prime \prime }\,,
\notag \\
R_{tm}^{tn} &=&R_{rm}^{rn}=-\dfrac{1}{2r}\,\left( f^{2}\right) ^{\prime
}\,\delta _{m}^{n}\,,  \label{RiemannBH} \\
R_{kl}^{mn} &=&\dfrac{1}{r^{2}}\,\left( k-f^{2}\right) \,\delta _{\lbrack
kl]}^{[mn]}\,,  \notag
\end{eqnarray}%
where prime denotes radial derivative. The Ricci tensor $R_{\nu }^{\mu
}=R_{\nu \lambda }^{\mu \lambda }$ has the components%
\begin{eqnarray}
R_{t}^{t} &=&R_{r}^{r}=-\dfrac{1}{2r}\,\left[ r\left( f^{2}\right) ^{\prime
\prime }+\left( D-2\right) \left( f^{2}\right) ^{\prime }\right] \,,  \notag
\\
R_{m}^{n} &=&-\dfrac{1}{r^{2}}\,\delta _{m}^{n}\,\left[ r\left( f^{2}\right)
^{\prime }+\left( D-3\right) \left( f^{2}-k\right) \right] \,,
\end{eqnarray}%
and the Ricci scalar $R=R_{\mu \nu }^{\mu \nu }$ is
\begin{equation}
R=-\dfrac{1}{r^{2}}\,\left[ r^{2}\left( f^{2}\right) ^{\prime \prime
}+2\left( D-2\right) r\left( f^{2}\right) ^{\prime }+\left( D-2\right)
\left( D-3\right) \left( f^{2}-k\right) \right] \,.
\end{equation}

For a static solution with a topology equal to the one of the transversal
section, we assume an ansatz for the gauge field in the form%
\begin{equation}
A_{\mu }=\phi \left( r\right) \,\delta _{\mu }^{t}\,,  \label{A}
\end{equation}%
with the associated field strength%
\begin{equation}
F_{\mu \nu }=E(r)\,\left( \delta _{\mu }^{t}\delta _{\nu }^{r}-\delta _{\nu
}^{t}\delta _{\mu }^{r}\right) \,,  \label{F_ansatz}
\end{equation}%
where the electric field is given by
\begin{equation}
E(r)=-\phi ^{\prime }(r)\,.  \label{def_E}
\end{equation}

We solve the electric potential in the static ansatz (\ref{BH}), (\ref%
{F_ansatz}), where\ $F^{2}=-2E^{2}$, using the only non-vanishing component
of the Maxwell-type equation (\ref{E_mu}),
\begin{equation}
\mathcal{E}^{t}=-\frac{d}{dr}\,\left( r^{D-2}E\left. \frac{d\mathcal{L}}{%
dF^{2}}\right\vert _{F^{2}=-2E^{2}}\right) =0\,,  \label{Et}
\end{equation}%
what leads to the generalized Gauss' law%
\begin{equation}
E\left. \frac{d\mathcal{L}}{dF^{2}}\right\vert _{F^{2}=-2E^{2}}=-\frac{q}{%
r^{D-2}}\,.  \label{r(E)}
\end{equation}%
Here, $q$ is an integration constant related to the electric charge. Notice
that the first integral of Eq.(\ref{r(E)}) does not depend explicitly on the
metric, but only on the function $E(r)$.\ The algebraic equation in $E$ can
be solved as long as the explicit form of NED action is given, and implies
that the electric field should vanish for $q=0$.

We define the electric potential at infinity measured with respect to the
event horizon $r_{+}$ as $\Phi =\phi (\infty )-\phi (r_{+})$.

On the other hand, integrating out Eq.(\ref{def_E}) one obtains the electric
potential at the distance $r$ measured with respect to radial infinity,%
\begin{equation}
\phi (r)=-\int\limits_{\infty }^{r}dv\,E(v)\,,  \label{potential}
\end{equation}%
such that the quantity of physical interest $\Phi $ is the potential
evaluated at the horizon,%
\begin{equation}
\Phi =-\phi (r_{+})\,.
\end{equation}

In order to solve the function $f^{2}(r)$ in the metric, we write the only
independent components of the Einstein and Lanczos tensors,%
\begin{eqnarray}
G_{t}^{t} &=&G_{r}^{r}=\frac{D-2}{2r^{2}}\,\left[ r\left( f^{2}\right)
^{\prime }+\left( D-3\right) \left( f^{2}-k\right) -\left( D-1\right) \frac{%
r^{2}}{\ell ^{2}}\right] \,,  \notag \\
H_{t}^{t} &=&H_{r}^{r}=\alpha \,\left( D-2\right) \left( D-3\right) \left(
D-4\right) \,\frac{k-f^{2}}{r^{3}}\,\left[ \left( f^{2}\right) ^{\prime
}-\left( D-5\right) \,\frac{k-f^{2}}{2r}\right] \,.  \label{components}
\end{eqnarray}

A necessary and sufficient condition on the NED Lagrangian density is the
Weak Energy Condition on the symmetric energy-momentum tensor%
\begin{equation}
T_{\mu \nu }\,u^{\mu }u^{\nu }\leq 0\,,  \label{WEC}
\end{equation}
that ensures that an observer measures a non-negative energy density $\rho
_{NED}=-T_{\mu \nu }\,u^{\mu }u^{\nu }$\ for a timelike vector $u^{\mu }$.
For charged static black holes, the electromagnetic stress tensor satisfies $%
T_{t}^{t}=T_{r}^{r}$, such that the weak energy condition is equivalent to%
\textbf{\ }%
\begin{equation}
T_{t}^{t}=T_{r}^{r}=\mathcal{L}+4E^{2}\frac{d\mathcal{L}}{dF^{2}}\geq 0\,,
\label{T components}
\end{equation}%
where the Lagrangian $\mathcal{L}$ and its derivatives are evaluated at $%
F^{2}=-2E^{2}$.

The above inequality restricts the function\textbf{\ }$\mathcal{L}$,\textbf{%
\ }but not its derivative.\ Indeed, in the asymptotic region the generalized
Gauss' law implies $E\,\frac{d\mathcal{L}}{dF^{2}}\simeq 0$ and, assuming
that electric field vanishes asymptotically, the weak energy condition leads
to $\mathcal{L}\geq 0$\ for large $r$. On\textbf{\ }the other hand, the
asymptotic behavior of $\frac{d\mathcal{L}}{dF^{2}}$\ remains arbitrary.
Indeed, for Maxwell electrodynamics and Born-Infeld-like Lagrangians, the
expression $\frac{d\mathcal{L}}{dF^{2}}$\ is finite for $r\rightarrow \infty
$. Also, for the Lagrangians of the type $(F^{2})^{p}$, the derivative
vanishes when $p>1$, and it is divergent if $p<1$. Additionally, one may
demand the finiteness of the total energy, that can be expressed as%
\begin{equation}
\int\limits_{0}^{\infty }dr\,r^{D-2}\,T_{r}^{r}(r)<\infty \,.
\label{finiteness tot E}
\end{equation}%
Note that the above requirement on the EM energy, applied to black hole
solutions, also includes the interior region protected by the horizon \cite%
{Aiello-Ferraro-GiribetHI}.

The equations of motion $\mathcal{E}_{t}^{t}=\mathcal{E}_{r}^{r}=0$ read%
\begin{eqnarray}
\frac{16\pi G\,r^{2}}{D-2}\,T_{r}^{r} &=&r\left( f^{2}\right) ^{\prime
}+\left( D-3\right) \left( f^{2}-k\right)  \notag \\
&&-\left( D-1\right) \,\frac{r^{2}}{\ell ^{2}}+2\alpha \,\left( D-3\right)
\left( D-4\right) \,\frac{k-f^{2}}{r}\,\left[ \left( f^{2}\right) ^{\prime
}-\left( D-5\right) \,\frac{k-f^{2}}{2r}\right] \,.  \label{diff_f}
\end{eqnarray}%
One can show, using Eqs.(\ref{Et}) and (\ref{diff_f}), that $\mathcal{E}%
_{n}^{m}=0$ is identically satisfied.

The differential equation (\ref{diff_f}) is integrable, because it can be
cast in the form%
\begin{equation}
\left[ r^{D-3}\left( f^{2}-k\right) \left( 1-\alpha \left( D-3\right) \left(
D-4\right) \frac{f^{2}-k}{r^{2}}\right) \right] ^{\prime }=\frac{D-1}{\ell
^{2}}\,r^{D-2}+\frac{16\pi G}{D-2}\,r^{D-2}T_{r}^{r}\,\,,  \label{total der}
\end{equation}%
what leads to the general solution%
\begin{equation}
\left( f^{2}-k\right) \left( 1-\alpha \left( D-3\right) \left( D-4\right) \,%
\frac{f^{2}-k}{r^{2}}\right) =\frac{r^{2}}{\ell ^{2}}-\frac{\mu }{r^{D-3}}+%
\frac{16\pi G\,\mathcal{T}(q,r)}{\left( D-2\right) r^{D-3}}\,,
\label{f quadratic}
\end{equation}%
where $\mu $ is an integration constant of dimension [mass$\,\times 16\pi G$%
], and the function $\mathcal{T}(q,r)$ for an arbitrary NED Lagrangian is
given by%
\begin{eqnarray}
\mathcal{T}(q,r) &=&\int\limits_{\infty }^{r}dv\,v^{D-2}\,T_{r}^{r}(v)
\notag \\
&=&\int\limits_{\infty }^{r}dv\,\left( \rule{0in}{0.18in}v^{D-2}\mathcal{L}%
(v)-4qE(v)\right)   \notag \\
&=&\frac{1}{D-1}\,\left. \left( \rule{0pt}{12pt}r^{D-1}\mathcal{L}%
-qrE+\left( D-2\right) 4q\phi \right) \right\vert _{\infty }^{r}\,.
\label{I(q,r)}
\end{eqnarray}%
The Gauss law (\ref{r(E)}) has been used to eliminate $d\mathcal{L}/dF^{2}$
from the integral, so that $\mathcal{T}$ depends on the integration constant
$q$. For a general procedure for Lovelock gravity coupled to NED see, e.g.,
\cite{Mazharimousavi-Gurtug-Halilsoy}.

Electromagnetism does not deform the asymptotic region since the relation $%
T(q,\infty )=0$\ is identically satisfied according to Eq.(\ref{I(q,r)}).

Then, the metric function in the static solution of EGB gravity coupled to
NED is obtained solving the quadratic equation (\ref{f quadratic}) in $f^{2}$%
. The existence of a real root is ensured by the condition%
\begin{equation}
\mathcal{T}(q,r)\leq \frac{\left( D-2\right) r^{D-1}}{16\pi G\,}\left( \frac{%
1}{4\alpha \left( D-3\right) \left( D-4\right) }-\frac{1}{\ell ^{2}}+\frac{%
\mu }{r^{D-1}}\right) \,,
\end{equation}%
that is proved to be satisfied for sufficiently large $r$, as the r.h.s. is
always positive (see the inequality in (Eq.\ref{l_eff pm})). Thus, the
metric possesses two branches,
\begin{equation}
f_{\pm }^{2}(r)=k+\frac{r^{2}}{2\alpha \left( D-3\right) \left( D-4\right) }%
\left[ 1\pm \sqrt{1-4\alpha \left( D-3\right) \left( D-4\right) \left( \frac{%
1}{\ell ^{2}}-\frac{\mu }{r^{D-1}}+\frac{16\pi G\,\mathcal{T}(q,r)}{\left(
D-2\right) r^{D-1}}\right) }\right] \,.  \label{2 metric functions}
\end{equation}

The ground state $\mu =0$, $q=0$ corresponds to two AdS vacua,%
\begin{equation}
f_{\pm }^{2}(r)_{vac}=k+\frac{r^{2}}{\ell _{e\!f\!f}^{(\pm )2}}\,.
\end{equation}%
However, it has been shown in \cite{Boulware-Deser} that the vacuum $%
f_{+}^{2}(r)_{vac}$ is unstable and the graviton has negative mass, while
the solution $f_{-}^{2}(r)_{vac}$ is stable and is free of ghosts. For a
general solution, from (\ref{2 metric functions}) in the weak limit of GB
coupling, we have
\begin{eqnarray}
f_{+}^{2}(r) &=&k+r^{2}\left( \frac{1}{\alpha \left( D-3\right) \left(
D-4\right) }-\frac{1}{\ell ^{2}}\right) +\frac{\mu }{r^{D-3}}-\frac{16\pi G\,%
\mathcal{T}(q,r)}{\left( D-2\right) r^{D-3}}+\mathcal{O}(\alpha )\,, \\
f_{-}^{2}(r) &=&k+\frac{r^{2}}{\ell ^{2}}-\frac{\mu }{r^{D-3}}+\frac{16\pi
G\,\mathcal{T}(q,r)}{\left( D-2\right) r^{D-3}}+\mathcal{O}(\alpha )\,,
\end{eqnarray}%
because $\mathcal{T}$ does not depend on the constant $\alpha $. The
opposite sign in the mass parameter $\mu $ in $f_{+}^{2}(r)$ indicates
instabilities of the graviton so that it is not of physical interest for our
discussion below.

On the other hand, the function $f_{-}^{2}(r)$ in the limit $\alpha
\rightarrow 0$ describes static black holes of Einstein-Hilbert AdS gravity
coupled to NED. Because of this reason, henceforth, we consider only the
negative branch of the metric, $f(r)\equiv f_{-}(r)$,
\begin{equation}
f^{2}(r)=k+\frac{r^{2}}{2\alpha \left( D-3\right) \left( D-4\right) }\left[
1-\sqrt{1-4\alpha \left( D-3\right) \left( D-4\right) \left( \frac{1}{\ell
^{2}}-\frac{\mu }{r^{D-1}}+\frac{16\pi G\,\mathcal{T}(q,r)}{\left(
D-2\right) r^{D-1}}\right) }\right] \,.  \label{general sol}
\end{equation}

When NED Lagrangian corresponds to the one of Maxwell electromagnetism $%
\mathcal{L}_{Maxwell}\left( F^{2}\right) =-F^{2}$, the function $\mathcal{T}%
(q,r)$\ in Eq.(\ref{general sol}) becomes $\mathcal{T}_{Maxwell}=\frac{2q^{2}%
}{\left( D-3\right) r^{D-3}}$, what reproduces the charged black hole
solution first found in \cite{Wiltshire}. Expanding\textbf{\ }$f^{2}$\ for
large $r$, one can notice that the electromagnetic part possesses the same
fall-off as in Reissner-Nordstrom case.

In general, the contribution of NED to $f^{2}$ is smaller than the one of
the mass term, and can therefore be neglected for large $r$. Indeed, using
Eq.(\ref{l_eff pm}),\textbf{\ }one can prove that, in the asymptotic region,
the metric function and its radial derivative behave as\textbf{\ }%
\begin{eqnarray}
f^{2} &=&k+\frac{r^{2}}{\ell _{e\!f\!f}^{2}}-\frac{\mu }{1-\frac{2\alpha }{%
\ell _{e\!f\!f}^{2}}\,\left( D-3\right) \left( D-4\right) }\frac{1}{r^{D-3}}+%
\mathcal{O}\left( \frac{1}{r^{2D-6}}\right) \,,  \label{f_asympt} \\
(f^{2})^{\prime } &=&\frac{2r}{\ell _{e\!f\!f}^{2}}+\frac{\left( D-3\right)
\mu }{1-\frac{2\alpha }{\ell _{e\!f\!f}^{2}}\,\left( D-3\right) \left(
D-4\right) }\frac{1}{r^{D-2}}+\mathcal{O}\left( \frac{1}{r^{2D-5}}\right) \,.
\label{df_asympt}
\end{eqnarray}%
This fact will make evident that the NED term\textbf{\ }$\mathcal{T}(q,r)$%
\textbf{\ }in Eq.(\ref{general sol}) does not produce additional
contributions to the energy of the system, as we shall discuss in Section %
\ref{Black hole mass}.

In absence of electromagnetic fields, we have that $\mathcal{T}(0,r)=0$,
what means that the solution (\ref{general sol}) reduces to the topological
version of Boulware-Deser black holes in AdS spaces \cite%
{Boulware-Deser,Wheeler,Cai}.

Different NED models have been proposed which possess particle-like
solutions whose both\ electromagnetic and gravitational fields are regular
everywhere. However, this does not imply that there are no curvature
singularities.

The interior of the black hole is described by the metric function obtained
from Eq.(\ref{total der}) as
\begin{eqnarray}
f_{\text{in}}^{2}(r) &=&k+\frac{r^{2}}{2\alpha \left( D-3\right) \left(
D-4\right) }\,\left[ \rule{0pt}{20pt}1\pm \right.   \notag \\
&&\pm \left. \sqrt{1-4\alpha \left( D-3\right) \left( D-4\right) \left(
\frac{1}{\ell ^{2}}-\frac{c}{r^{D-1}}+\frac{16\pi
G\int_{0}^{r}dv\,v^{D-2}\,T_{r}^{r}(v)}{\left( D-2\right) r^{D-1}}\right) }%
\right] \,,
\end{eqnarray}%
where $c$\ is the integration constant. In consequence, when one imposes the
finiteness condition on the energy-momentum tensor at the origin,%
\begin{equation}
\lim_{r\rightarrow 0}\frac{1}{r^{D-1}}\int\limits_{0}^{r}dv\,v^{D-2}%
\,T_{r}^{r}(v)<\infty \,,
\end{equation}%
the metric function takes the value $f_{\text{in}}^{2}(0)=k\pm \sqrt{\frac{c%
}{\alpha \left( D-3\right) \left( D-4\right) r^{D-5}}}$. For $c\neq 0$, this
is finite only in five dimensions, otherwise\textbf{\ }$c$ must vanish%
\textbf{.} Further analysis is needed to relate $c$ to the asymptotic mass
parameter\textbf{\ }$\mu $\textbf{,} what would imply new conditions in
order to remove the conical singularity at the origin. One may also demand $%
\mathcal{L}$\ to be single-valued, continuous and differentiable. For a more
detailed discussion on these issues for particular cases see, e.g., Refs.%
\cite{Diaz-Rubiera,Aiello-Ferraro-GiribetHI}.

So far, we have seen that for any nonlinear electrodynamics theory coupled
to EGB AdS gravity, both the metric (\ref{general sol}) and the electric
potential (\ref{potential}) can be determined from the explicit form of the
Lagrangian $\mathcal{L}(F^{2})$. We illustrate this with a few examples in
the next section.

\section{Charged black holes in particular NED theories}

\subsection{Born-Infeld electrodynamics}

Born-Infeld electrodynamics \cite{Born-Infeld} is described by the
Lagrangian density%
\begin{equation}
\mathcal{L}_{BI}\left( F^{2}\right) =4b^{2}\left( 1-\sqrt{1+\frac{F^{2}}{%
2b^{2}}}\right) \,,
\end{equation}%
where the coupling parameter $b$ (with dimension of mass) is related to the
string tension $\alpha ^{\prime }$ as $b=1/2\pi \alpha ^{\prime }$. This
Lagrangian reduces to the Maxwell case in the weak-coupling limit $%
b\rightarrow \infty $. Generally speaking, when a density $\mathcal{L}(F^{2})
$ recovers the Maxwell theory in weak-coupling limit, i.e., $\mathcal{L}%
(F^{2})=-F^{2}+\mathcal{O}\left( 1/b^{2}\right) $, it is said to be
Born-Infeld-type.

The BI energy-momentum tensor has the form%
\begin{equation}
T_{\nu }^{\mu }=4b^{2}\delta _{\nu }^{\mu }\left( 1-\sqrt{1+\frac{F^{2}}{%
2b^{2}}}\right) +\dfrac{4F^{\mu \lambda }F_{\nu \lambda }}{\sqrt{1+\frac{%
F^{2}}{2b^{2}}}}\,,
\end{equation}%
and it generates the electric field%
\begin{equation}
E(r)=\frac{q}{\sqrt{\frac{q^{2}}{b^{2}}+r^{2D-4}}}\,.  \label{E(r)}
\end{equation}%
The corresponding electric potential is given by the formula (\ref{potential}%
). Performing a variable change in the integral, $u=(r/v)^{2D-4}$, it can be
expressed in terms of the hypergeometric function $\mathcal{F}(q,r)=\left.
_{2}F_{1}\right. \left( \frac{1}{2},\frac{D-3}{2D-4};\frac{3D-7}{2D-4};-%
\frac{q^{2}}{b^{2}r^{2D-4}}\right) $ (see Appendix \ref{Hyper}), and the
solution for the potential is%
\begin{equation}
\phi (r)=\frac{q}{\left( D-3\right) r^{D-3}}\,\mathcal{F}(q,r)\,.
\label{solution_phi}
\end{equation}%
Then, the integration constant $\Phi =-\phi (r_{+})$ reads%
\begin{equation}
\Phi =-\frac{q}{\left( D-3\right) r_{+}^{D-3}}\,\mathcal{F}\left(
q,r_{+}\right) \,.  \label{phi}
\end{equation}

In order to find the metric for the black hole with Born-Infeld electric
charge, we solve explicitly the integral (\ref{I(q,r)}) as
\begin{equation}
\mathcal{T}_{BI}(q,r)=\frac{4b^{2}r^{D-1}}{D-1}\left( 1-\sqrt{1+\frac{q^{2}}{%
b^{2}r^{2D-4}}}\right) +\frac{4\left( D-2\right) q^{2}}{\left( D-1\right)
\left( D-3\right) r^{D-3}}\,\mathcal{F}(q,r)\,,
\end{equation}%
and replacing in Eq.(\ref{general sol}), we obtain
\begin{eqnarray}
f^{2}(r) &=&k+\frac{r^{2}}{2\alpha \left( D-3\right) \left( D-4\right) }%
\left\{ 1-\left[ 1-4\alpha \left( D-3\right) \left( D-4\right) \left( \rule%
{0in}{0.26in}\frac{1}{\ell ^{2}}-\frac{\mu }{r^{D-1}}\right. \right. \right.
\notag \\
&&+\left. \left. \left. \frac{64\pi G\,b^{2}}{\left( D-1\right) \left(
D-2\right) }\left( 1-\sqrt{1+\frac{q^{2}}{b^{2}r^{2D-4}}}\right) +\frac{%
64\pi G\,q^{2}\mathcal{F}(q,r)}{\left( D-1\right) \left( D-3\right) r^{2D-4}}%
\,\right) \right] ^{1/2}\right\} \,.  \label{BI metric}
\end{eqnarray}%
This class of black holes has been discussed in Ref.\cite{Wiltshire}. The
generalization to non-Abelian gauge fields has been studied in Ref.\cite%
{Bostani-Farhangkhah}. In the limit of vanishing GB coupling, the metric
reduces to the one of topological Einstein-BI black holes in AdS spaces \cite%
{Fernando-Krug,Dey,Cai-Pang-Wang}.

\subsection{Conformally invariant electrodynamics}

Born-Infeld Lagrangian in higher dimensions is a physically sensible
extension of four-dimensional Maxwell electrodynamics. However, if one is
interested in a generalization of the conformal invariance property of 4D
Maxwell theory, there exist NED actions given as power-law functions of the
form%
\begin{equation}
\mathcal{L}_{CED}\left( F^{2}\right) =-2\chi \,F^{2p},  \label{MH}
\end{equation}%
where $\chi $ is a positive coupling constant \cite{Hassaine-Martinez}. Then
the conformal invariance $g_{\mu \nu }\rightarrow \Omega ^{2}g_{\mu \nu }$, $%
A_{\mu }\rightarrow A_{\mu }$ is realized for the power $p=D/4$.

The energy-momentum tensor for $A_{\mu }$ reads%
\begin{equation}
T_{\nu }^{\mu }=-2\chi \left( \delta _{\nu }^{\mu }-4p\,\frac{F^{\mu \lambda
}F_{\nu \lambda }}{F^{2}}\right) F^{2p}\,,
\end{equation}%
and it produces the electric field%
\begin{equation}
E(r)=\frac{\tilde{q}}{r^{\beta }}\,,
\end{equation}%
where $\beta =\frac{D-2}{2p-1}$ and $\tilde{q}=\left( \frac{\left( -1\right)
^{p+1}q}{2^{p}\,p\chi }\right) ^{\frac{\beta }{D-2}}$. When one demands
conformal invariance ($p=D/4$), the electric field takes the 4D Maxwell's
form, $E=\tilde{q}/r^{2}$, in any dimension.

Then, one can calculate explicitly the function (\ref{I(q,r)}) in the metric,%
\begin{equation}
\mathcal{T}_{CED}(q,r)=-\frac{2\left( D-2\right) \left( -2\right) ^{p}\tilde{%
q}^{2p}\chi }{\beta \left( \beta -1\right) }\frac{1}{r^{\beta -1}}\,,
\end{equation}%
that, plugged in Eq.(\ref{general sol}), produces a line element which
matches the form of the black holes found in Ref.\cite%
{Maeda-Hassaine-Martinez} for EGB AdS gravity.

\subsection{Logarithmic electrodynamics}

NED Lagrangians that contain logarithmic terms in the electromagnetic field
strength appear in the description of vacuum polarization effects. These
terms were obtained as exact 1-loop corrections for electrons in a uniform
electromagnetic field background by Euler and Heisenberg \cite%
{Heisenberg-Euler}, and therefore are a typical feature of quantum
electrodynamics effective actions.

Furthermore, logarithmic ED Lagrangians come as a realization of the old
idea of removing singularities in the gravitational field, in a similar way
as the BI electrodynamics removes divergences in the electric field. They
have also been used to describe an equation of state of radiation in an
alternative mechanism for inflation \cite{Altshuler}.

A simple example of a BI-like Lagrangian with a logarithmic term, that can
be added as a correction to the original BI one, was discussed in Ref.\cite%
{Soleng} in asymptotically flat Einstein gravity in $D=4$. This model does
not cancel the curvature singularity for small $r$, but makes the
Kretschmann invariant behave as $1/r^{4}$, which is a weaker singularity
than in, e.g., Schwarzschild or Reissner-Nordstr\"{o}m black holes.

In an arbitrary dimension, the logarithmic ED lagrangian has the form%
\begin{equation}
\mathcal{L}_{Log}(F^{2})=-8b^{2}\,\ln \left( 1+\frac{F^{2}}{8b^{2}}\right)
\,.
\end{equation}%
It can be shown from Eq.(\ref{r(E)}) that the electric field has two
branches, but only one features the Maxwell limit ($b\rightarrow \infty $),%
\begin{equation}
E(r)=\frac{2b^{2}}{q}\,\left( r^{D-2}-\sqrt{r^{2D-4}+\frac{q^{2}}{b^{2}}}%
\right) \,.
\end{equation}%
Considering this, the electric potential reads%
\begin{equation}
\phi (r)=-\frac{2b^{2}r^{D-1}}{q\left( D-1\right) }\left( 1-\sqrt{1+\frac{%
q^{2}}{b^{2}r^{2D-4}}}\right) -\frac{2q\left( D-2\right) \,\mathcal{F}(q,r)}{%
\left( D-1\right) \left( D-3\right) r^{D-3}}\,,
\end{equation}%
where $\mathcal{F}(q,r)=\left. _{2}F_{1}\right. \left( \frac{1}{2},\frac{D-3%
}{2D-4};\frac{3D-7}{2D-4};-\frac{q^{2}}{b^{2}r^{2D-4}}\right) $.

The electromagnetic contribution to the metric is then given by
\begin{eqnarray}
\mathcal{T}_{Log}(q,r) &=&-\frac{8b^{2}r^{D-1}}{D-1}\,\ln \left[ \frac{%
2b^{2}r^{D-2}}{q^{2}}\,\left( \sqrt{r^{2D-4}+\frac{q^{2}}{b^{2}}}%
-r^{D-2}\right) \right]   \notag \\
&&+\frac{8b^{2}\left( 2D-3\right) r}{\left( D-1\right) ^{2}}\,\left( r^{D-2}-%
\sqrt{r^{2D-4}+\frac{q^{2}}{b^{2}}}\right) +\frac{8q^{2}\left( D-2\right)
^{2}\mathcal{F}(q,r)}{\left( D-1\right) ^{2}\left( D-3\right) \,r^{D-3}}\,.
\end{eqnarray}%
Using the fact that $\mathcal{F}(0,r)=1=\mathcal{F}(q,\infty )$, one can
show explicitly that electromagnetism vanishes asymptotically, that is, $%
\mathcal{T}_{Log}(q,\infty )=0$.\ Also, in the zero charge limit, $\mathcal{T%
}_{Log}(0,r)$ vanishes, as expected.

It is straightforward to write down the metric function $f^{2}(r)$ by
plugging in the above expression into EGB metric in Eq.(\ref{general sol}).
This general solution reduces to the one of Ref.\cite{Soleng} in 4D Einstein
gravity without cosmological constant.

A more realistic version of logarithmic NED action is given by the
Hoffmann-Infeld model \cite{Hoffmann-Infeld}, that do remove singularities
in both gravitational and electric fields for static solutions. This theory
is described by the Lagrangian%
\begin{equation}
\mathcal{L}_{HI}=4b^{2}\left( \rule{0in}{0.18in}1-\eta (F^{2})-\log \eta
(F^{2})\right) \,,
\end{equation}%
where $\eta (F^{2})=-\frac{F^{2}}{4b^{2}}\left( 1-\sqrt{1+\frac{F^{2}}{2b^{2}%
}}\right) ^{-1}$. It can be easily checked from the expansion $\eta
(F^{2})=1+\frac{F^{2}}{8b^{2}}+\mathcal{O}(1/b^{4})$ that $\mathcal{L}%
_{HI}(F^{2})$ is also a BI-like Lagrangian in the weak-coupling limit. In $%
D=5$, a solution to this model was discussed in Ref.\cite%
{Aiello-Ferraro-GiribetHI}.

\section{Variational principle and boundary terms}

Any gravity theory is not defined only by its equations of motion in the
bulk, but also by the set of boundary conditions that guarantees that the
action is truly stationary. In general, this implies that the original bulk
action must be supplemented by a boundary term $\beta $,%
\begin{equation}
\tilde{I}=I_{0}+\int\limits_{\partial \mathcal{M}}d^{D-1}x\,\beta \,,
\end{equation}%
such that the problem of a well-posed action principle reduces to the
on-shell cancelation of the total surface term of the theory, that is,
\begin{equation}
\delta \tilde{I}=\int\limits_{\partial \mathcal{M}}d^{D-1}x\,\left( \Theta
_{0}+\delta \beta \right) =0\,.
\end{equation}%
In our case, the term $\Theta _{0}$ is given as the sum of Eqs.(\ref{BT_EGB}%
) and (\ref{BT_matter}) and, in principle, $\beta $ can be split in two
parts, namely, $\beta =\beta _{grav}+\beta _{NED}$.

A gravitational action whose variation vanishes for a Dirichlet condition on
the metric requires the addition of (generalized) Gibbons-Hawking terms.
This is particularly easy to see in Gauss-normal coordinates
\begin{equation}
ds^{2}=g_{\mu \nu }\,dx^{\mu }dx^{\nu }=N^{2}\left( r\right)
\,dr^{2}+h_{ij}(r,x)\,dx^{i}dx^{j}\,.  \label{radial foliation}
\end{equation}%
We will consider a manifold with a single boundary $\partial \mathcal{M}$ at
$r=\infty $, parameterized by the coordinates $x^{i}$, and such that $h_{ij}$
is the induced metric on it. The extrinsic properties of the boundary are
given in terms of the outward-pointing normal $n_{\mu }=\left(
n_{r},n_{i}\right) =(N,\vec{0})$. In particular, we define the extrinsic
curvature as the Lie derivative of the induced metric along this normal,
\begin{equation}
K_{ij}=-\frac{1}{2}\,\pounds _{n}h_{ij}=-\frac{1}{2N}\,h_{ij}^{\prime }\,.
\end{equation}%
As it is written in Appendix \ref{Gauss-normal}, different components of the
Christoffel symbol can be expressed in terms of the extrinsic curvature. In
doing so, the surface term $\Theta _{0}=\Theta _{grav}+\Theta _{NED}$ has
the form%
\begin{eqnarray}
\Theta _{0} &=&\frac{1\,}{8\pi G\left( D-2\right) \left( D-3\right) }\,\sqrt{%
-h}\,\delta _{\lbrack ii_{1}i_{2}]}^{[jj_{1}j_{2}]}\,\left[ \frac{1}{2}%
\,\left( h^{-1}\delta h\right) _{k}^{i}K_{j}^{k}+\delta K_{j}^{i}\right]
\times   \notag \\
&&\qquad \times \left[ \delta _{j_{1}}^{i_{1}}\delta
_{j_{2}}^{i_{2}}+2\alpha \left( D-2\right) \left( D-3\right) \left( \frac{1}{%
2}\,\mathcal{R}_{j_{1}j_{2}}^{i_{1}i_{2}}-K_{j_{1}}^{i_{1}}K_{j_{2}}^{i_{2}}%
\right) \right]   \notag \\
&&+4\sqrt{-h}\,\frac{d\mathcal{L}}{dF^{2}}\,NF^{ri}\delta A_{i}\,,
\label{Theta projected}
\end{eqnarray}%
where the determinant of the metric satisfies$\ \sqrt{-g}=N\sqrt{-h}$ and $%
\mathcal{R}_{kl}^{ij}(h)$ is the intrinsic curvature of the boundary, which
is related to the spacetime Riemann tensor by $R_{kl}^{ij}=\mathcal{R}%
_{kl}^{ij}-K_{k}^{i}K_{l}^{j}+K_{l}^{i}K_{k}^{j}$ (see Appendix \ref%
{Gauss-normal}).

In order to cancel $\Theta _{NED}$ part of the surface term, it is a
sufficient condition to take $\delta A_{i}=0$ at $\partial \mathcal{M}$.
This means that $\beta $ does not depend on the electromagnetic field, i.e.,
$\beta =\beta _{grav}$. On the other hand, there is a systematic
construction of generalized Gibbons-Hawking terms for Gauss-Bonnet and, in
general, any Lovelock theory \cite{Myers, Miskovic-OleaDCG}, which for the
present case gives%
\begin{equation}
\beta =\frac{\sqrt{-h}}{8\pi G\left( D-2\right) \left( D-3\right) }\,\delta
_{\lbrack i_{1}i_{2}i_{3}]}^{[j_{1}j_{2}j_{3}]}\,K_{j_{1}}^{i_{1}}\left[
\delta _{j_{2}}^{i_{2}}\delta _{j_{3}}^{i_{3}}+2\alpha \left( D-2\right)
\left( D-3\right) \,\left( \frac{1}{2}\,\mathcal{R}%
_{j_{2}j_{3}}^{i_{2}i_{3}}-\frac{1}{3}\,K_{j_{2}}^{i_{2}}K_{j_{3}}^{i_{3}}%
\right) \right] \,,  \label{gen GH}
\end{equation}%
or, in the form which is commonly found in the literature,%
\begin{equation*}
\beta =\frac{\sqrt{-h}}{8\pi G}\,\left[ K+2\alpha \,\left( K\left(
K^{ij}K_{ij}-\frac{1}{3}K^{2}\right) -\frac{2}{3}%
\,K_{k}^{i}K_{j}^{k}K_{i}^{j}-2\mathcal{G}^{ij}K_{ij}\right) \right] \,,
\end{equation*}%
where $\mathcal{G}^{ij}=\mathcal{R}^{ij}-\frac{1}{2}\,\mathcal{R\,}h^{ij}$\
is the Einstein tensor associated to the boundary metric.

In doing so, the corresponding Dirichlet variation of the action is
\begin{eqnarray}
\Theta _{0}+\delta \beta &=&\frac{\sqrt{-h}}{16\pi G\left( D-3\right) \left(
D-4\right) }\,\delta _{\lbrack
i\,i_{1}i_{2}i_{3}]}^{[j\,j_{1}j_{2}j_{3}]}(h^{-1}\delta
h)_{j}^{i}K_{j_{1}}^{i_{1}}\,\left[ \rule{0in}{0.22in}\delta
_{j_{2}}^{i_{2}}\delta _{j_{3}}^{i_{3}}\right. +  \notag \\
&&+\left. 2\alpha \left( D-3\right) \left( D-4\right) \,\left( \frac{1}{2}\,%
\mathcal{R}_{j_{2}j_{3}}^{i_{2}i_{3}}-\frac{1}{3}%
\,K_{j_{2}}^{i_{2}}K_{j_{3}}^{i_{3}}\right) \right] +4\,\sqrt{-h}\,\frac{d%
\mathcal{L}}{dF^{2}}\,NF^{ri}\delta A_{i}\,.
\end{eqnarray}

Notice that for a radial foliation of the spacetime, the on-shell variation
of the action can be cast in the form%
\begin{equation}
\delta \tilde{I}=\int\limits_{\partial \mathcal{M}}d^{D-1}x\sqrt{-h}%
\,\left( \frac{1}{2}\,\pi ^{ij}\delta h_{ij}+\pi ^{i}\delta A_{i}\right) \,,
\end{equation}%
where $\pi ^{ij}$ and $\pi ^{i}$ are the canonical momenta conjugate to $%
h_{ij}$ and $A_{i}$, respectively. If one uses $\pi ^{ij}$ as a
quasilocal-stress tensor in AdS gravity, the conserved quantities derived
from it are divergent in the asymptotic region. In other words, a well-posed
variational principle is not necessarily linked to the problem of finiteness
of the charges and action.

In the context of AdS/CFT correspondence, the standard way to deal with the
regularization problem in a background-independent way is the addition of
local counterterms at the boundary, which are constructed using holographic
normalization. However, the inclusion of higher-curvature terms in the
action turns this procedure considerably more complicated. A practical
method to circumvent this obstacle in EGB-AdS gravity is to assume the same
form of the counterterms as in the EH case, but with arbitrary coefficients
\cite{Brihaye-Radu, Astefanesei et al, Liu-Sabra}. The coefficients are then
fixed requiring the convergence of the action for particular solutions of
the theory. It is clear from this construction that the series cannot be
obtained for an arbitrary dimension.

The fact that in AdS gravity the leading-order of the asymptotic expansion
of the extrinsic curvature is proportional to the one of the boundary metric
opens the possibility to consider counterterms which depends on the
extrinsic curvature, as well. In this alternative scheme (known as
Kounterterm regularization), the boundary terms are related to either
topological invariants or Chern-Simons forms in the corresponding
dimensions. In this way, it is possible to skip the technicalities of
holographic procedures and to write down a general expression for them in
any dimension,%
\begin{equation}
I=I_{0}+c_{D-1}\int\limits_{\partial \mathcal{M}}d^{D-1}x\,B_{D-1}\,,
\label{I}
\end{equation}%
where $c_{D-1}$ is a given constant. For EH AdS gravity, the Kounterterm
series was shown in Refs.\cite{Olea-K,Olea-KerrBH} as a given polynomial of
the extrinsic and intrinsic curvatures, which defines a well-posed action
principle. In general, the action (\ref{I}) varies as%
\begin{equation}
\delta I=\int\limits_{\partial \mathcal{M}}d^{D-1}x\,\Theta
=\int\limits_{\partial \mathcal{M}}d^{D-1}x\,\left( \Theta _{grav}+\Theta
_{NED}+c_{D-1}\,\delta B_{D-1}\right) \,,  \label{var_I}
\end{equation}%
such that the boundary term in (\ref{I}) makes the action to have an
extremum on-shell and solves the regularization problem, as well.

For a given dimension, the series $B_{D-1}$ possesses the remarkable
property of preserving its form for EGB-AdS gravity \cite{Kofinas-Olea-GB}
and, in general, any theory of the Lovelock type \cite{Kofinas-Olea}. In
what follows, we use the explicit form of the boundary terms to construct
the general variation of the action in tensorial notation.

\subsection{Even dimensions ($D=2n$)}

In even dimensions $D=2n>4$, the boundary term $B_{2n-1}$ in (\ref{I}) is
given by the $n$-th Chern form \cite{Kofinas-Olea-GB}
\begin{eqnarray}
B_{2n-1} &=&2n\sqrt{-h}\int\limits_{0}^{1}dt\,\delta _{\lbrack i_{1}\cdots
i_{2n-1}]}^{[j_{1}\cdots j_{2n-1}]}\,K_{j_{1}}^{i_{1}}\left( \frac{1}{2}\,%
\mathcal{R}%
_{j_{2}j_{3}}^{i_{2}i_{3}}-t^{2}K_{j_{2}}^{i_{2}}K_{j_{3}}^{i_{3}}\right)
\times   \notag \\
&&\qquad \cdots \times \left( \frac{1}{2}\,\mathcal{R}%
_{j_{2n-2}j_{2n-1}}^{i_{2n-2}i_{2n-1}}-t^{2}K_{j_{2n-2}}^{i_{2n-2}}K_{j_{2n-1}}^{i_{2n-1}}\right) \,,
\label{B-even}
\end{eqnarray}%
that is the scalar density whose derivative is locally equivalent to the
Euler invariant (globally they differ by the Euler characteristic of the
manifold, $\chi (\mathcal{M})$). The integration in the continuous parameter
$t$ generates the coefficients when the boundary term is expanded as a
polynomial. The constant $c_{2n-1}$ in front of the boundary term $B_{2n-1}$
which produces a well-defined variational principle is given in terms of the
effective AdS radius as%
\begin{equation}
c_{2n-1}=-\frac{1}{16\pi G}\frac{(-\ell _{e\!f\!f}^{2})^{n-1}}{n\left(
2n-2\right) !}\left( 1-\frac{2\alpha }{\ell _{e\!f\!f}^{2}}\,\left(
2n-2\right) \left( 2n-3\right) \right) \,.  \label{c2n-1}
\end{equation}%
It can been proven that the same choice of $c_{2n-1}$ ensures the
convergence of the Euclidean action. The total surface term $\Theta $ can be
read off from the on-shell variation of the action (\ref{var_I}), that in
this case is%
\begin{eqnarray}
\delta I_{2n} &=&\frac{1}{16\pi G\left( 2n-2\right) !2^{n-1}}%
\int\limits_{\partial \mathcal{M}}d^{2n-1}x\sqrt{-h}\,\delta _{\lbrack
i_{1}\cdots i_{2n-1}]}^{[j_{1}\cdots j_{2n-1}]}\,\left[ \left( h^{-1}\delta
h\right) _{k}^{i_{1}}K_{j_{1}}^{k}+2\delta K_{j_{1}}^{i_{1}}\right] \times
\notag \\
&&\times \left[ \rule{0in}{0.24in}\left( \delta _{\lbrack
j_{2}j_{3}]}^{[i_{2}i_{3}]}+2\alpha \left( 2n-2\right) \left( 2n-3\right)
R_{j_{2}j_{3}}^{i_{2}i_{3}}\right) \delta _{\lbrack
j_{4}j_{5}]}^{[i_{4}i_{5}]}\cdots \delta _{\lbrack
j_{2n-2}j_{2n-1}]}^{[i_{2n-2}i_{2n-1}]}\right.   \notag \\
&&\left. \rule{0in}{0.24in}-\left( -\ell _{e\!f\!f}^{2}\right) ^{n-1}\left(
1-\frac{2\alpha }{\ell _{e\!f\!f}^{2}}\,\left( 2n-2\right) \left(
2n-3\right) \,\right) R_{j_{2}j_{3}}^{i_{2}i_{3}}\cdots
R_{j_{2n-2}j_{2n-1}}^{i_{2n-2}i_{2n-1}}\right]   \notag \\
&&+4\,\int\limits_{\partial \mathcal{M}}d^{2n-1}x\sqrt{-h}\,\frac{d\mathcal{%
L}}{dF^{2}}\,NF^{ri}\delta A_{i}\,.  \label{varI-2n}
\end{eqnarray}%
The reader can easily check that imposing the asymptotically locally AdS
condition for the spacetime, i.e.,
\begin{equation}
R_{\mu \nu }^{\alpha \beta }+\frac{1}{\ell _{e\!f\!f}^{2}}\,\delta _{\lbrack
\mu \nu ]}^{[\alpha \beta ]}=0\,,\qquad \text{at }\partial \mathcal{M}\,,
\label{AAdS}
\end{equation}%
identically cancels the leading-order divergences in the gravitational part
of the above variation. As a remarkable feature of the addition of
Kounterterms, all other divergent terms in (\ref{varI-2n}) are exactly
cancelled out. In this way, the finite contribution is coupled to the
conformal metric that is kept fixed at the boundary. The NED part of the
surface term vanishes for a Dirichlet boundary condition for the transversal
components of $A_{\mu }$,
\begin{equation}
\delta A_{i}=0\,,\qquad \text{at }\partial \mathcal{M}\,.  \label{bc A}
\end{equation}

\subsection{Odd dimensions ($D=2n+1$)}

The extrinsic regularization developed for odd-dimensional Einstein-Hilbert
AdS gravity \cite{Olea-K} can be mimicked for EGB AdS theory, just replacing
the AdS radius $\ell $ by the effective one $\ell _{e\!f\!f}$ in the
boundary terms. Thus, the Kounterterms series is given in terms of the
parametric integrations
\begin{eqnarray}
B_{2n} &=&2n\sqrt{-h}\int\limits_{0}^{1}dt\int\limits_{0}^{t}ds\,\delta
_{\lbrack i_{1}\cdots i_{2n}]}^{[j_{1}\cdots
j_{2n}]}\,K_{j_{1}}^{i_{1}}\delta _{j_{2}}^{i_{2}}\left( \frac{1}{2}\,%
\mathcal{R}%
_{j_{3}j_{4}}^{i_{3}i_{4}}-t^{2}K_{j_{3}}^{i_{3}}K_{j_{4}}^{i_{4}}+\frac{%
s^{2}}{\ell _{e\!f\!f}^{2}}\,\delta _{j_{3}}^{i_{3}}\delta
_{j_{4}}^{i_{4}}\right) \times  \notag \\
&&\cdots \times \left( \frac{1}{2}\,\mathcal{R}%
_{j_{2n-1}j_{2n}}^{i_{2n-1}i_{2n}}-t^{2}K_{j_{2n-1}}^{i_{2n-1}}K_{j_{2n}}^{i_{2n}}+%
\frac{s^{2}}{\ell _{e\!f\!f}^{2}}\,\delta _{j_{2n-1}}^{i_{2n-1}}\delta
_{j_{2n}}^{i_{2n}}\right) \,.  \label{B_2n}
\end{eqnarray}%
The corresponding constant for this case incorporates the information of the
theory through the GB coupling in the form
\begin{eqnarray}
c_{2n} &=&-\frac{1}{16\pi G}\frac{(-\ell _{e\!f\!f}^{2})^{n-1}}{n\left(
2n-1\right) !}\left( 1-\frac{2\alpha \left( 2n-1\right) \left( 2n-2\right) }{%
\ell _{e\!f\!f}^{2}}\right) \left[ \int\limits_{0}^{1}dt\,\left(
1-t^{2}\right) ^{n-1}\right] ^{-1}  \notag \\
&=&-\frac{1}{16\pi G}\frac{2(-\ell _{e\!f\!f}^{2})^{n-1}}{n\left(
2n-1\right) !\beta (n,\frac{1}{2})}\left( 1-\frac{2\alpha \left( 2n-1\right)
\left( 2n-2\right) }{\ell _{e\!f\!f}^{2}}\right) \,,  \label{c2n}
\end{eqnarray}%
where $\beta (n,\frac{1}{2})=\frac{2^{2n-1}\left( n-1\right) !^{2}}{\left(
2n-1\right) !}$ is the Beta function for those arguments.

The total action varies on-shell as
\begin{eqnarray}
\delta I_{2n+1} &=&\frac{1}{2^{n-1}16\pi G\,(2n-1)!}\int\limits_{\partial
\mathcal{M}}d^{2n}x\,\sqrt{-h}\,\delta _{\lbrack i_{1}\cdots
i_{2n}]}^{[j_{1}\cdots j_{2n}]}\,\left[ \left( h^{-1}\delta h\right)
_{k}^{i_{1}}K_{j_{1}}^{k}+2\delta K_{j_{1}}^{i_{1}}\right] \delta
_{j_{2}}^{i_{2}}\times   \notag \\
&&\times \left[ \rule{0in}{0.26in}\left( \delta _{\lbrack
j_{3}j_{4}]}^{[i_{3}i_{4}]}+2\alpha \left( 2n-1\right) \left( 2n-2\right)
\,R_{j_{3}j_{4}}^{i_{3}i_{4}}\right) \delta _{\lbrack
j_{5}j_{6}]}^{[i_{5}i_{6}]}\cdots \delta _{\lbrack
j_{2n-1}j_{2n}]}^{[i_{2n-1}i_{2n}]}\right.   \notag \\
&&+\left. 16\pi G\,(2n-1)!nc_{2n}\int\limits_{0}^{1}dt\,\left(
R_{j_{3}j_{4}}^{i_{3}i_{4}}+\frac{t^{2}}{\ell _{e\!f\!f}^{2}}\,\delta
_{\lbrack j_{3}j_{4}]}^{[i_{3}i_{4}]}\right) \cdots \left(
R_{j_{2n-1}j_{2n}}^{i_{2n-1}i_{2n}}+\frac{t^{2}}{\ell _{e\!f\!f}^{2}}%
\,\delta _{\lbrack j_{2n-1}j_{2n}]}^{[i_{2n-1}i_{2n}]}\right) \right]
\notag \\
&&+nc_{2n}\int\limits_{\partial \mathcal{M}}d^{2n}x\,\sqrt{-h}%
\,\int\limits_{0}^{1}dt\,t\,\delta _{\lbrack i_{1}\cdots
i_{2n}]}^{[j_{1}\cdots j_{2n}]}\!\left[ \left( h^{-1}\delta h\right)
_{k}^{i_{1}}\left( K_{j_{1}}^{k}\delta _{j_{2}}^{i_{2}}-\delta
_{j_{1}}^{k}K_{j_{2}}^{i_{2}}\right) +2\delta _{j_{2}}^{i_{2}}\delta
K_{j_{2}}^{i_{2}}\right] \times   \notag \\
&&\times \left( \frac{1}{2}\,\mathcal{R}%
_{j_{3}j_{4}}^{i_{3}i_{4}}-t^{2}K_{j_{3}}^{i_{3}}K_{j_{4}}^{i_{4}}+\frac{%
t^{2}}{\ell _{e\!f\!f}^{2}}\,\delta _{j_{3}}^{i_{3}}\delta
_{j_{4}}^{i_{4}}\right) \cdots \left( \frac{1}{2}\,\mathcal{R}%
_{j_{2n-1}j_{2n}}^{i_{2n-1}i_{2n}}-t^{2}K_{j_{2n-1}}^{i_{2n-1}}K_{j_{2n}}^{i_{2n}}+%
\frac{t^{2}}{\ell _{e\!f\!f}^{2}}\,\delta _{j_{2n-1}}^{i_{2n-1}}\delta
_{j_{2n}}^{i_{2n}}\right)   \notag \\
&&+4\,\int\limits_{\partial \mathcal{M}}d^{2n}x\sqrt{-h}\,\frac{d\mathcal{L}%
}{dF^{2}}\,NF^{ri}\delta A_{i}\,.  \label{varI_2n+1}
\end{eqnarray}%
Then, the surface term from the electromagnetic part vanishes when fixing
the gauge potential at the boundary, Eq.(\ref{bc A}).

Checking explicitly the cancellation of the leading-order divergences in the
above action proves to be slightly more complicated than in the
even-dimensional case, but one may reason as follows: the second and third
lines cancel out when taking the condition on the asymptotic curvature (\ref%
{AAdS}) for the particular value of $c_{2n}$ given by Eq.(\ref{c2n}). On the
other hand, for any asymptotically AdS spacetime, the extrinsic curvature $%
K_{j}^{i}$ has a regular expansion in the asymptotic region, $K_{j}^{i}=%
\frac{1}{\ell _{e\!f\!f}}\,\delta _{j}^{i}+\mathcal{O}\left( 1/r\right) $.
This means that variations of the extrinsic curvature vanishes at the
leading-order in the vicinity of $\partial \mathcal{M}$. These conditions
guarantee a well-posed action principle for odd-dimensional EGB AdS gravity,
issue that was discussed previously in Ref.\cite{Kofinas-Olea-GB}.

\section{Conserved quantities}

\subsection{Electric charge}

We will first derive the electric charge $Q$ as a conserved quantity
associated to $U(1)$ gauge symmetry $\delta _{\lambda }A_{\mu }=\partial
_{\mu }\lambda $, $\delta _{\lambda }g_{\mu \nu }=0$, as its computation
does not depend on the spacetime dimension. The gravitational part of the
surface term in Eq.(\ref{var_I}) is gauge-invariant, such that it implies
the conservation of the Noether current
\begin{equation}
\delta _{\lambda }I=\int\limits_{\mathcal{M}}d^{D}x\,\partial _{\mu }J^{\mu
}(\lambda )=4\int\limits_{\mathcal{M}}d^{D}x\,\partial _{\mu }\left( \sqrt{%
-g}\,F^{\mu \nu }\frac{d\mathcal{L}}{dF^{2}}\,\partial _{\nu }\lambda
\right) \text{\thinspace },
\end{equation}%
where the current $J^{\mu }$\ transforms as a vector density of weight $+1$.
In the radial foliation (\ref{radial foliation}), the electric charge is
then the normal component of the above current%
\begin{equation}
Q\left[ \lambda \right] =\int\limits_{\partial \mathcal{M}}d^{D-1}x\,\frac{1%
}{N}\,n_{\mu }J^{\mu }(\lambda )\,,  \label{Q(lambda)}
\end{equation}%
which, using the fact that $n_{\mu }$ is covariantly constant, can be
rewritten as
\begin{equation}
\sqrt{-h}\,n_{\mu }\frac{d\mathcal{L}}{dF^{2}}\,F^{\mu \nu }\partial _{\nu
}\lambda =\partial _{i}\left( \lambda \sqrt{-h}\,n_{\mu }F^{\mu i}\,\frac{d%
\mathcal{L}}{dF^{2}}\right) -\lambda n_{\mu }\sqrt{-h}\,\mathcal{E}^{\mu }\,.
\end{equation}%
As a consequence, since $\mathcal{E}^{\mu }=0$, we are able to write down
the integrand in Eq.(\ref{Q(lambda)}) as a total derivative. In order to use
the Stokes' theorem we take a timelike ADM foliation for the line element on
$\partial \mathcal{M}$ with the coordinates $x^{i}=\left( t,y^{m}\right) $,
as
\begin{equation}
h_{ij}\,dx^{i}dx^{j}=-\tilde{N}^{2}(t)dt^{2}+\sigma _{mn}(dy^{m}+\tilde{N}%
^{m}dt)(dy^{n}+\tilde{N}^{n}dt)\,,\qquad \sqrt{-h}=\tilde{N}\sqrt{\sigma }\,,
\label{ADMboundary}
\end{equation}%
that is generated by the timelike normal vector $u_{i}=(u_{t},u_{m})=(-%
\tilde{N},\vec{0})$. The metric $\sigma _{mn}$ describes the geometry of the
boundary of spatial section at constant time $\Sigma _{\infty }$.

Setting $\lambda =1$, the $U(1)$ charge reads
\begin{equation}
Q=4\int\limits_{\Sigma _{\infty }}d^{D-2}y\sqrt{\sigma }\,u_{i}\,NF^{ri}%
\frac{d\mathcal{L}}{dF^{2}}\,.
\end{equation}%
For the static black hole metric (\ref{BH}) (where $N=1/f^{2}$ and $\tilde{N}%
=f^{2}$) and the electromagnetic field strength (\ref{F_ansatz}), one
obtains a general formula for NED electric charge%
\begin{equation}
Q=-4\text{Vol}(\Gamma _{D-2})\,\lim_{r\rightarrow \infty }\,\left( r^{D-2}E\,%
\frac{d\mathcal{L}}{dF^{2}}\right) \,.  \label{Qelectric}
\end{equation}%
Finally, using the generalized Gauss law (\ref{r(E)}), it is possible to
define a finite electric charge of the black hole
\begin{equation}
Q=4\text{Vol}(\Gamma _{D-2})\,q\,,  \label{charge}
\end{equation}%
for an arbitrary NED Lagrangian. However, this definition does not guarantee
only by itself that the electric field is well-behaved in the asymptotic
region.

\subsection{Black hole mass \ \label{Black hole mass}}

In order to calculate the conserved quantities associated to global
isometries of the spacetime, we first consider the action of diffeomorphisms
$\delta x^{\mu }=\xi ^{\mu }(x)$ on the fields $g_{\mu \nu }$ and $A_{\mu }$
in terms of the Lie derivative,
\begin{eqnarray}
\delta _{\xi }g_{\mu \nu } &=&\pounds _{\xi }g_{\mu \nu }\equiv -\left(
\nabla _{\mu }\xi _{\nu }+\nabla _{\nu }\xi _{\mu }\right) \,,  \notag \\
\delta _{\xi }A_{\mu } &=&\pounds _{\xi }A_{\mu }\equiv -\partial _{\mu
}\left( \xi ^{\nu }A_{\nu }\right) +\xi ^{\nu }F_{\mu \nu }\,,
\end{eqnarray}%
what implies the transformation rule of the Christoffel symbol%
\begin{equation}
\pounds _{\xi }\Gamma _{\mu \nu }^{\alpha }=\frac{1}{2}\,\left( R_{\ \mu \nu
\beta }^{\alpha }+R_{\ \nu \mu \beta }^{\alpha }\right) \xi ^{\beta }-\frac{1%
}{2}\,\left( \nabla _{\mu }\nabla _{\nu }\xi ^{\alpha }+\nabla _{\nu }\nabla
_{\mu }\xi ^{\alpha }\right) .
\end{equation}

This leads to the transformation of the volume element, Jacobian and
Lagrangian density $\mathcal{L}_{0}$ defined by Eq.(\ref{bulk action}) as
\begin{eqnarray}
\delta _{\xi }\left( d^{D}x\right) &=&d^{D}x\,\partial _{\mu }\xi ^{\mu }\,,
\notag \\
\delta _{\xi }\sqrt{-g} &=&-\sqrt{-g}\,\nabla _{\mu }\xi ^{\mu }\,, \\
\delta _{\xi }\mathcal{L}_{0} &=&\frac{\partial \mathcal{L}_{0}}{\partial
g_{\mu \nu }}\,\pounds _{\xi }g_{\mu \nu }+\frac{\partial \mathcal{L}_{0}}{%
\partial \Gamma _{\mu \nu }^{\beta }}\,\pounds _{\xi }\Gamma _{\mu \nu
}^{\beta }+\frac{\partial \mathcal{L}_{0}}{\partial A_{\mu }}\,\pounds _{\xi
}A_{\mu }+\xi ^{\mu }\partial _{\mu }\mathcal{L}_{0}\mathcal{\,}.  \notag
\end{eqnarray}%
Then the total action (\ref{I}) transforms under diffeomorphisms as%
\begin{eqnarray}
\delta _{\xi }I &=&\int\limits_{\mathcal{M}}d^{D}x\,\left[ \pounds _{\xi
}\left( \sqrt{-g}\,\mathcal{L}_{0}\right) +\partial _{\mu }\left( \sqrt{-g}%
\,\xi ^{\mu }\,\mathcal{L}_{0}\right) \right] +c_{D-1}\int\limits_{\partial
\mathcal{M}}d^{D-1}x\,\left[ \pounds _{\xi }B_{D-1}+\partial _{i}\left( \xi
^{i}B_{D-1}\right) \right]  \notag \\
&=&\int\limits_{\partial \mathcal{M}}d^{D-1}x\,n_{\mu }\left( \frac{1}{N}%
\,\Theta ^{\mu }(\xi )+\sqrt{-h}\,\xi ^{\mu }\mathcal{L}_{0}+c_{D-1}\,n^{\mu
}\,\partial _{i}\left( \xi ^{i}B_{D-1}\right) \right) +  \notag \\
&&-\int\limits_{\mathcal{M}}d^{D}x\,\sqrt{-g}\left( \frac{1}{16\pi G}\,%
\mathcal{E}^{\mu \nu }\pounds _{\xi }g_{\mu \nu }+4\,\mathcal{E}^{\mu }%
\pounds _{\xi }A_{\mu }\right) \,,  \label{diff_Ibulk}
\end{eqnarray}
where $\Theta (\xi )=\frac{1}{N}\,n_{\mu }\Theta ^{\mu }(\xi )$ is the
surface term in Eq.(\ref{var_I}) evaluated in the corresponding Lie
derivative of the fields.

The Noether current derived from the diffeormorphic invariance, $\delta
_{\xi }I=\int_{\mathcal{M}}d^{D}x\,\partial _{\mu }J^{\mu }(\xi )=0$ is,
therefore,%
\begin{equation}
J^{\mu }(\xi )=\Theta ^{\mu }(\xi )+\sqrt{-g}\,\xi ^{\mu }\mathcal{L}%
_{0}+c_{D-1}Nn^{\mu }\,\partial _{i}\left( \xi ^{i}B_{D-1}\right) \mathcal{\,%
}.  \label{J^mu}
\end{equation}%
The conservation law $\partial _{\mu }J^{\mu }=0$ implies the existence of a
conserved quantity, which corresponds to the normal component of the current
$J^{\mu }$,%
\begin{equation}
Q\left[ \xi \right] =\int\limits_{\partial \mathcal{M}}d^{D-1}x\,\frac{1}{N}%
\,n_{\mu }J^{\mu }(\xi )\,.  \label{Qxi1}
\end{equation}%
In general, it is not guaranteed that the Noether charge can be written as
surface integral in $(D-2)$ dimensions. However, for the action $I$, the
radial component $J^{r}=\frac{1}{N}\,n_{\mu }J^{\mu }\,$\ in the foliation (%
\ref{radial foliation}) is globally a total derivative on $\partial \mathcal{%
M}$, i.e.,
\begin{equation}
\,J^{r}=\partial _{j}\left( \sqrt{-h}\,\xi ^{i}\,\left(
q_{i}^{j}+q_{(0)i}^{j}\right) \right) \,.
\end{equation}%
The splitting in the above integrand is justified as follows: $q_{i}^{j}$
produces the mass and other conserved quantities for black hole solutions.
As we will show below, this part of the charge identically vanishes for the
vacuum states of the theory. The term $q_{(0)i}^{j}$ gives rise to a vacuum
energy, which is present only in odd dimensions.

Therefore, the conserved charges $Q[\xi ]$ of the theory for a given set of
asymptotic Killing vectors $\{\xi \}$ are expressed as integrals on $\Sigma
_{\infty }$ (whose metric has been defined in Eq.(\ref{ADMboundary})),%
\begin{equation}
Q[\xi ]=\int\limits_{\Sigma _{\infty }}d^{D-2}y\,\sqrt{\sigma }\,u_{j}\,\xi
^{i}\,\left( q_{i}^{j}+q_{(0)i}^{j}\right) \,.  \label{Qxigeneral}
\end{equation}

\subsubsection{Even dimensions \ \label{Even mass}}

In even dimensions, the expression for the surface term $\Theta (\xi )$ is
obtained from (\ref{varI-2n}) by replacing the variations by the
corresponding Lie derivative of the fields,%
\begin{eqnarray}
\frac{1}{N}\,n_{\mu }\Theta ^{\mu }(\xi ) &=&\frac{\sqrt{-h}}{16\pi G\left(
2n-2\right) !2^{n-1}}\,\delta _{\lbrack i_{1}\cdots i_{2n-1}]}^{[j_{1}\cdots
j_{2n-1}]}\,\left[ \left( h^{-1}\pounds _{\xi }h\right)
_{k}^{i_{1}}K_{j_{1}}^{k}+2\pounds _{\xi }K_{j_{1}}^{i_{1}}\right] \times
\notag \\
&&\times \left[ \rule{0in}{0.24in}\left( \delta _{\lbrack
j_{2}j_{3}]}^{[i_{2}i_{3}]}+2\alpha \left( 2n-2\right) \left( 2n-3\right)
R_{j_{2}j_{3}}^{i_{2}i_{3}}\right) \delta _{\lbrack
j_{4}j_{5}]}^{[i_{4}i_{5}]}\cdots \delta _{\lbrack
j_{2n-2}j_{2n-1}]}^{[i_{2n-2}i_{2n-1}]}\right.  \notag \\
&&\left. \rule{0in}{0.24in}-\left( -\ell _{e\!f\!f}^{2}\right) ^{n-1}\left(
1-\frac{2\alpha }{\ell _{e\!f\!f}^{2}}\,\left( 2n-2\right) \left(
2n-3\right) \,\right) R_{j_{2}j_{3}}^{i_{2}i_{3}}\cdots
R_{j_{2n-2}j_{2n-1}}^{i_{2n-2}i_{2n-1}}\right]  \notag \\
&&+4\,\sqrt{-h}\,\frac{d\mathcal{L}}{dF^{2}}\,NF^{ri}\pounds _{\xi }A_{i}\,.
\label{vecTheta2n}
\end{eqnarray}

As a result of the Noether procedure, the integrand in the conserved charge (%
\ref{Qxigeneral}) is
\begin{eqnarray}
q_{i}^{j} &=&\frac{1}{16\pi G\left( 2n-2\right) !2^{n-2}}\,\delta _{\lbrack
i_{1}\cdots i_{2n-1}]}^{[jj_{2}\cdots j_{2n-1}]}\,K_{i}^{i_{1}}\times  \notag
\\
&&\times \left[ \rule{0in}{0.24in}\left( \delta _{\lbrack
j_{2}j_{3}]}^{[i_{2}i_{3}]}+2\alpha \left( 2n-2\right) \left( 2n-3\right)
R_{j_{2}j_{3}}^{i_{2}i_{3}}\right) \delta _{\lbrack
j_{4}j_{5}]}^{[i_{4}i_{5}]}\cdots \delta _{\lbrack
j_{2n-2}j_{2n-1}]}^{[i_{2n-2}i_{2n-1}]}\right.  \notag \\
&&\left. \rule{0in}{0.24in}-\left( -\ell _{e\!f\!f}^{2}\right) ^{n-1}\left(
1-\frac{2\alpha }{\ell _{e\!f\!f}^{2}}\,\left( 2n-2\right) \left(
2n-3\right) \,\right) R_{j_{2}j_{3}}^{i_{2}i_{3}}\cdots
R_{j_{2n-2}j_{2n-1}}^{i_{2n-2}i_{2n-1}}\right] \,,  \label{q(2n)}
\end{eqnarray}%
plus a NED contribution due to the last line in Eq.(\ref{vecTheta2n}), what
vanishes for black hole solutions, as shown below. At the same time, $%
q_{(0)i}^{j}=0$ for even dimensions.

The second and third lines in the expression (\ref{q(2n)}) can be seen as a
polynomial of rank $(n-1)$ in the Riemann tensor and the Kronecker delta $%
\frac{1}{\ell _{e\!f\!f}^{2}}\delta _{\lbrack j_{2}j_{3}]}^{[i_{2}i_{3}]}$,
which can be factorized by $\left( R_{j_{2}j_{3}}^{i_{2}i_{3}}+\frac{1}{\ell
_{e\!f\!f}^{2}}\,\delta _{\lbrack j_{2}j_{3}]}^{[i_{2}i_{3}]}\right) $.\ As
a consequence of the fact that for any maximally symmetric spacetime this
factor vanishes, any conserved quantity defined on it will be identically
zero in even dimensions.

The energy of black hole solution to EGB AdS gravity coupled to NED (\ref{BH}%
) is computed evaluating the formula (\ref{Qxigeneral}) for the Killing
vector $\xi ^{i}=(1,\vec{0})$ and the unit normal $u_{i}=(-f,\vec{0})$ which
defines a constant-time slice,
\begin{eqnarray}
M &\equiv &Q\left[ \partial _{t}\right] =-\frac{1}{16\pi G\left( 2n-2\right)
!2^{n-2}}\int\limits_{\Gamma _{2n-2}}d^{2n-2}\varphi \sqrt{\gamma }%
\,f\,r^{2n-2}\,\delta _{\lbrack n_{1}\cdots n_{2n-2}]}^{[m_{1}\cdots
m_{2n-2}]}\,K_{t}^{t}\times   \notag \\
&&\times \left[ \rule{0in}{0.24in}\left( \delta _{\lbrack
m_{1}m_{2}]}^{[n_{1}n_{2}]}+2\alpha \left( 2n-2\right) \left( 2n-3\right)
R_{m_{1}m_{2}}^{n_{1}n_{2}}\right) \delta _{\lbrack
m_{3}m_{4}]}^{[n_{3}n_{4}]}\cdots \delta _{\lbrack
m_{2n-3}m_{2n-2}]}^{[n_{2n-3}n_{2n-2}]}\right.   \notag \\
&&\left. \rule{0in}{0.24in}-\left( -\ell _{e\!f\!f}^{2}\right) ^{n-1}\left(
1-\frac{2\alpha }{\ell _{e\!f\!f}^{2}}\,\left( 2n-2\right) \left(
2n-3\right) \,\right) R_{m_{1}m_{2}}^{n_{1}n_{2}}\cdots
R_{m_{2n-3}m_{2n-2}}^{n_{2n-3}n_{2n-2}}\right] \,.
\end{eqnarray}%
From the explicit form of the extrinsic curvature%
\begin{equation}
K_{j}^{i}=-\frac{1}{2N}\,h^{ik}h_{kj}^{\prime }=\left(
\begin{array}{cc}
-f^{\prime } & 0 \\
0 & -\frac{f}{r}\,\delta _{n}^{m}%
\end{array}%
\right) \,,  \label{KBH}
\end{equation}%
and the boundary components of the Riemann tensor in Eq.(\ref{RiemannBH}),
one obtains a general formula for the mass in even dimensions,
\begin{eqnarray}
M &=&\frac{\text{Vol}(\Gamma _{2n-2})}{16\pi G}\,\lim_{r\rightarrow \infty
}\,r^{2n-2}(f^{2})^{\prime }\left[ 1-2\alpha \left( 2n-2\right) \left(
2n-3\right) \,\frac{f^{2}-k}{r^{2}}-\right.   \notag \\
&&\left. -\left( 1-\frac{2\alpha }{\ell _{e\!f\!f}^{2}}\,\left( 2n-2\right)
\left( 2n-3\right) \,\right) \ell _{e\!f\!f}^{2n-2}\left( \frac{f^{2}-k}{%
r^{2}}\right) ^{n-1}\right] \,.  \label{Noether_2n}
\end{eqnarray}%
In order to relate the above expression to the integration constant $\mu $,
one must consider the asymptotic expansion of the metric function\textbf{\ }(%
\ref{f_asympt}) in the following way,
\begin{eqnarray}
\frac{f^{2}-k}{r^{2}} &=&\frac{1}{\ell _{e\!f\!f}^{2}}-\frac{\mu }{1-\frac{%
2\alpha }{\ell _{e\!f\!f}^{2}}\,\left( 2n-3\right) \left( 2n-4\right) }\frac{%
1}{r^{2n-1}}+\mathcal{O}\left( \frac{1}{r^{4n-4}}\right) \,,  \label{f-k} \\
\left( \frac{f^{2}-k}{r^{2}}\right) ^{n-1} &=&\frac{1}{\ell _{e\!f\!f}^{2n-2}%
}-\frac{\left( n-1\right) \mu }{1-\frac{2\alpha }{\ell _{e\!f\!f}^{2}}%
\,\left( 2n-3\right) \left( 2n-4\right) }\frac{1}{\ell
_{e\!f\!f}^{2n-4}r^{2n-1}}+\mathcal{O}\left( \frac{1}{r^{4n-4}}\right) \,,
\end{eqnarray}%
and its derivative (\ref{df_asympt}). When expanded, Eq.(\ref{Noether_2n})
might contain divergences of order $r^{2n-1}$. It is then a remarkable fact
that the divergent terms cancel out for the particular value of $c_{2n-1}$
in Eq.(\ref{c2n-1}), what leaves a finite result for the energy
\begin{equation}
M=\frac{\left( 2n-2\right) \text{Vol}(\Gamma _{2n-2})\,\mu }{16\pi G}\,,
\label{M}
\end{equation}
in agreement with the expression found in, e.g., Ref.\cite{Deser-Tekin}.

Now we turn our attention to the NED contribution to the diffeomorphic
transformation of the action, that is, the last line in Eq.(\ref{vecTheta2n}%
). This part of the surface term produces, by virtue of the Noether theorem,
an additional piece with respect to the charge formula given by Eq.(\ref%
{q(2n)}), which is written in any dimension as%
\begin{equation}
Q_{NED}\left[ \xi \right] =-4\int\limits_{\Sigma _{\infty }}d^{D-2}y\,\sqrt{%
\sigma }\,u_{j}\,\frac{d\mathcal{L}}{dF^{2}}\,NF^{rj}\left( \xi
^{i}A_{i}\right) \,.  \label{QNED}
\end{equation}%
However, when we evaluate Eq.(\ref{QNED}) for the Killing vector $\xi
=\partial _{t}$ and the static black hole metric, we notice that
\begin{equation}
Q_{NED}\left[ \partial _{t}\right] =-4q\,\text{Vol}(\Gamma _{D-2})\,\phi
(\infty )=0\,,
\end{equation}%
as anticipated in the discussion following the deduction of the charge
formula.

\subsubsection{Odd dimensions}

The form of the surface term $\Theta (\xi )$ in odd dimensions ($D=2n+1$)
follows from the on-shell variation of the action, Eq.(\ref{varI_2n+1}). Its
expression is slightly more complicated than in the even-dimensional case
\begin{eqnarray}
\frac{1}{N}\,n_{\mu }\Theta ^{\mu }(\xi ) &=&\frac{\sqrt{-h}}{16\pi G\left(
2n-1\right) !2^{n-1}}\,\delta _{\lbrack i_{1}\cdots i_{2n}]}^{[j_{1}\cdots
j_{2n}]}\,\left[ \left( h^{-1}\pounds _{\xi }h\right)
_{k}^{i_{1}}K_{j_{1}}^{k}+2\pounds _{\xi }K_{j_{1}}^{i_{1}}\right] \,\delta
_{j_{2}}^{i_{2}}\times   \notag \\
&&\times \left[ \rule{0in}{0.26in}\left( \delta _{\lbrack
j_{3}j_{4}]}^{[i_{3}i_{4}]}+2\alpha \left( 2n-1\right) \left( 2n-2\right)
\,R_{j_{3}j_{4}}^{i_{3}i_{4}}\right) \delta _{\lbrack
j_{5}j_{6}]}^{[i_{5}i_{6}]}\cdots \delta _{\lbrack
j_{2n-1}j_{2n}]}^{[i_{2n-1}i_{2n}]}\right.   \notag \\
&&+\left. 16\pi G\,(2n-1)!nc_{2n}\int\limits_{0}^{1}dt\,\left(
R_{j_{3}j_{4}}^{i_{3}i_{4}}+\frac{t^{2}}{\ell _{e\!f\!f}^{2}}\,\delta
_{\lbrack j_{3}j_{4}]}^{[i_{3}i_{4}]}\right) \cdots \left(
R_{j2n-1j_{2n}}^{i_{2n-1}i_{2n}}+\frac{t^{2}}{\ell _{e\!f\!f}^{2}}\,\delta
_{\lbrack j_{2n-1}j_{2n}]}^{[i_{2n-1}i_{2n}]}\right) \right]   \notag \\
&&+nc_{2n}\sqrt{-h}\,\int\limits_{0}^{1}dt\,t\,\delta _{\lbrack i_{1}\cdots
i_{2n}]}^{[j_{1}\cdots j_{2n}]}\!\left[ \left( h^{-1}\delta h\right)
_{k}^{i_{1}}\left( K_{j_{1}}^{k}\delta _{j_{2}}^{i_{2}}-\delta
_{j_{1}}^{k}K_{j_{2}}^{i_{2}}\right) +2\delta _{j_{2}}^{i_{2}}\delta
K_{j_{2}}^{i_{2}}\right] \times   \notag \\
&&\times \left( \frac{1}{2}\,\mathcal{R}%
_{j_{3}j_{4}}^{i_{3}i_{4}}-t^{2}K_{j_{3}}^{i_{3}}K_{j_{4}}^{i_{4}}+\frac{%
t^{2}}{\ell _{e\!f\!f}^{2}}\,\delta _{j_{3}}^{i_{3}}\delta
_{j_{4}}^{i_{4}}\right) \cdots \left( \frac{1}{2}\,\mathcal{R}%
_{j_{2n-1}j_{2n}}^{i_{2n-1}i_{2n}}-t^{2}K_{j_{2n-1}}^{i_{2n-1}}K_{j_{2n}}^{i_{2n}}+%
\frac{t^{2}}{\ell _{e\!f\!f}^{2}}\,\delta _{j_{2n-1}}^{i_{2n-1}}\delta
_{j_{2n}}^{i_{2n}}\right)   \notag \\
&&+4\sqrt{-h}\,\frac{d\mathcal{L}}{dF^{2}}\,NF^{ri}\pounds _{\xi }A_{i}\,,
\end{eqnarray}%
where, for shortness' sake, we have chosen not to use the explicit form of $%
c_{2n}$ given by Eq.(\ref{c2n}).

In odd dimensions, the Noether charge appears as the sum of two parts, since
$q_{(0)i}^{j}$ in Eq.(\ref{Qxigeneral}) is no longer vanishing. The first
part takes the form%
\begin{eqnarray}
q_{i}^{j} &=&\frac{1}{16\pi G\left( 2n-1\right) !2^{n-2}}\,\delta _{\lbrack
i_{1}\cdots i_{2n}]}^{[jj_{2}\cdots j_{2n}]}\,K_{i}^{i_{1}}\delta
_{j_{2}}^{i_{2}}\times   \notag \\
&&\times \left[ \rule{0in}{0.26in}\left( \delta _{\lbrack
j_{3}j_{4}]}^{[i_{3}i_{4}]}+2\alpha \left( 2n-1\right) \left( 2n-2\right)
\,R_{j_{3}j_{4}}^{i_{3}i_{4}}\right) \delta _{\lbrack
j_{5}j_{6}]}^{[i_{5}i_{6}]}\cdots \delta _{\lbrack
j_{2n-1}j_{2n}]}^{[i_{2n-1}i_{2n}]}\right.   \notag \\
&&+\left. 16\pi G\left( 2n-1\right) !\,nc_{2n}\int\limits_{0}^{1}dt\left(
R_{j_{3}j_{4}}^{i_{3}i_{4}}+\frac{t^{2}}{\ell _{e\!f\!f}^{2}}\,\delta
_{\lbrack j_{3}j_{4}]}^{[i_{3}i_{4}]}\right) \cdots \left(
R_{j_{2n-1}j_{2n}}^{i_{2n-1}i_{2n}}+\frac{t^{2}}{\ell _{e\!f\!f}^{2}}%
\,\delta _{\lbrack j_{2n-1}j_{2n}]}^{[i_{2n-1}i_{2n}]}\right) \right]
\label{q(2n-1)}
\end{eqnarray}%
whereas the second one is given by%
\begin{eqnarray}
q_{(0)i}^{j} &=&nc_{2n}\,\int\limits_{0}^{1}dt\,t\,\delta _{\lbrack
ki_{2}\cdots i_{2n}]}^{[jj_{2}\cdots j_{2n}]}\left( K_{i}^{k}\delta
_{j_{2}}^{i_{2}}+K_{j_{2}}^{k}\delta _{i}^{i_{2}}\right) \left( \frac{1}{2}\,%
\mathcal{R}%
_{j_{3}j_{4}}^{i_{3}i_{4}}-t^{2}K_{j_{3}}^{i_{3}}K_{j_{4}}^{i_{4}}+\frac{%
t^{2}}{\ell _{e\!f\!f}^{2}}\,\delta _{j_{3}}^{i_{3}}\delta
_{j_{4}}^{i_{4}}\right) \times \cdots   \notag \\
&&\qquad \qquad \cdots \times \left( \frac{1}{2}\,\mathcal{R}%
_{j_{2n-1}j_{2n}}^{i_{2n-1}i_{2n}}-t^{2}K_{j_{2n-1}}^{i_{2n-1}}K_{j_{2n}}^{i_{2n}}+%
\frac{t^{2}}{\ell _{e\!f\!f}^{2}}\,\delta _{j_{2n-1}}^{i_{2n-1}}\delta
_{j_{2n}}^{i_{2n}}\right) \,.  \label{qzero}
\end{eqnarray}

We recall the fact that the constant $c_{2n}$ was chosen to cancel at least
the leading-order divergence in the variation of the action (\ref{varI_2n+1}%
). Thus, it can be readily checked that $q_{i}^{j}$ is identically zero for
global AdS spacetime which satisfies (\ref{LAdSleff}) in the bulk. This
means that the second and third lines in the expression (\ref{q(2n)}) are
again a polynomial of rank $(n-1)$ in the Riemann tensor and the Kronecker
delta $\frac{1}{\ell _{e\!f\!f}^{2}}\,\delta _{\lbrack
j_{2}j_{3}]}^{[i_{2}i_{3}]}$, where $R_{j_{2}j_{3}}^{i_{2}i_{3}}=-\frac{1}{%
\ell _{e\!f\!f}^{2}}\,\delta _{\lbrack j_{2}j_{3}]}^{[i_{2}i_{3}]}$ is a
root of it. Therefore, any maximally symmetric spacetime will have vanishing
mass and angular momentum due to the fact that $q_{i}^{j}=0$, such that all
the contributions to the vacuum energy will come necessarily from Eq.(\ref%
{qzero}), as shown below. On the other hand, the presence of $c_{2n}$ in the
formula of vacuum energy reflects the fact that its existence is entirely
due to the addition of the Kounterterm series (\ref{B_2n}).

Proceeding as in the even-dimensional case, we compute the black hole mass
evaluating the first term in the formula (\ref{Qxigeneral}),
\begin{eqnarray*}
M &=&\int\limits_{\Sigma _{\infty }}d^{D-2}y\,\sqrt{\sigma }\,u_{t}\,\xi
^{t}\,q_{t}^{t} \\
&=&-\frac{1}{16\pi G\left( 2n-1\right) !\,2^{n-2}}\,\lim_{r\rightarrow
\infty }\int\limits_{\Gamma _{2n-2}}d^{2n-2}\varphi \sqrt{\gamma }%
\,f\,r^{2n-1}\,\delta _{\lbrack n_{1}\cdots n_{2n-1}]}^{[m_{1}\cdots
m_{2n-1}]}\,K_{t}^{t}\delta _{m_{1}}^{n_{1}}\times \\
&&\times \left[ \rule{0in}{0.24in}\left( \delta _{\lbrack
m_{2}m_{3}]}^{[n_{2}n_{3}]}+2\alpha \left( 2n-1\right) \left( 2n-2\right)
R_{m_{2}m_{3}}^{n_{2}n_{3}}\right) \delta _{\lbrack
m_{4}m_{5}]}^{[n_{4}n_{5}]}\cdots \delta _{\lbrack
m_{2n-2}m_{2n-1}]}^{[n_{2n-2}n_{2n-1}]}\right. \\
&&+\left. 16\pi G\,(2n-1)!\,nc_{2n}\int\limits_{0}^{1}dt\,\left(
R_{m_{2}m_{3}}^{n_{2}n_{3}}+\frac{t^{2}}{\ell _{e\!f\!f}^{2}}\,\delta
_{\lbrack m_{2}m_{3}]}^{[n_{2}n_{3}]}\right) \cdots \left(
R_{m_{2n-2}m_{2n-1}}^{n_{2n-2}n_{2n-1}}+\frac{t^{2}}{\ell _{e\!f\!f}^{2}}%
\,\delta _{\lbrack m_{2n-2}m_{2n-1}]}^{[n_{2n-2}n_{2n-1}]}\right) \right] .
\end{eqnarray*}%
Using the Riemann tensor in Eq.(\ref{RiemannBH}) and the extrinsic curvature
for the generic black hole metric given by Eq.(\ref{KBH}), the above formula
reduces to%
\begin{eqnarray}
M &=&\frac{\text{Vol}(\Gamma _{2n-1})}{16\pi G}\,\lim_{r\rightarrow \infty
}\,r^{2n-1}(f^{2})^{\prime }\left[ 1-2\alpha \left( 2n-1\right) \left(
2n-2\right) \,\frac{f^{2}-k}{r^{2}}+\right.  \notag \\
&&+\left. 16\pi G\,(2n-1)!\,nc_{2n}\int\limits_{0}^{1}dt\,\left( \frac{%
k-f^{2}}{r^{2}}+\frac{t^{2}}{\ell _{e\!f\!f}^{2}}\right) ^{n-1}\right] \,.
\label{Noether2n+1}
\end{eqnarray}%
It is straightforward to express the mass $M$ in terms of the constant $\mu $
in the metric, by means of the expansion of the metric function\textbf{\ }in
Eq.(\ref{f-k}), its derivative (\ref{df_asympt}) and the last line in the
above relation,%
\begin{eqnarray}
\int\limits_{0}^{1}dt\,\left( \frac{k-f^{2}}{r^{2}}+\frac{t^{2}}{\ell
_{e\!f\!f}^{2}}\right) ^{n-1} &=&-\frac{1}{16\pi G\,(2n-1)!\,nc_{2n}}\left(
1-\frac{2\alpha }{\ell _{e\!f\!f}^{2}}\,\left( 2n-1\right) \left(
2n-2\right) \right) \times  \notag \\
&&\times \left( 1-\frac{\ell _{e\!f\!f}^{2}}{2}\frac{\left( 2n-1\right)
\,\mu }{1-\frac{4\alpha \left( 2n-2\right) \left( 2n-3\right) }{\ell
_{e\!f\!f}^{2}}}\,\frac{1}{r^{2n}}\right) +\mathcal{O}\left( \frac{1}{%
r^{4n-3}}\right) \,.
\end{eqnarray}%
Unless the constant $c_{2n}$ is fixed as in Eq.(\ref{c2n}), the formula (\ref%
{Noether2n+1}) contains divergences of order $r^{2n}$. Therefore, the
boundary term $c_{2n}B_{2n}$ plays a double role: it cancels out the
divergences in the Noether charge, but also contributes with a finite piece
to give the correct result for the mass
\begin{equation}
M=\frac{\left( 2n-1\right) \text{Vol}(\Gamma _{2n-1})\,\mu }{16\pi G}\,,
\end{equation}%
what matches the one in Ref.\cite{Deser-Tekin}.
In turn, the vacuum energy for AAdS black holes is reflected in the formula (%
\ref{qzero}), that in the black hole ansatz (\ref{BH}) adopts the form%
\begin{eqnarray}
E_{vac} &=&\int\limits_{\Sigma _{\infty }}d^{D-2}y\,\sqrt{\sigma }%
\,u_{t}\,\xi ^{t}\,q_{(0)t}^{t}  \notag \\
&=&2nc_{2n}\,\lim_{r\rightarrow \infty }\int\limits_{\Gamma
_{2n-1}}d^{2n-1}\varphi \sqrt{\gamma }\,r^{2n-1}f\,\delta _{\lbrack
n_{1}\cdots n_{2n-1}]}^{[m_{1}\cdots m_{2n-1}]}\,\left( K_{t}^{t}\,\delta
_{m_{1}}^{n_{1}}-K_{m_{1}}^{n_{1}}\right) \times \\
&&\int\limits_{0}^{1}dt\,t\,\left( \frac{1}{2}\,\mathcal{R}%
_{m_{2}m_{3}}^{n_{2}n_{3}}-t^{2}K_{m_{2}}^{n_{2}}K_{m_{3}}^{n_{3}}+\frac{%
t^{2}}{\ell ^{2}}\,\delta _{m_{2}}^{n_{2}}\delta _{m_{3}}^{n_{3}}\right)
\times \cdots  \notag \\
&&\qquad \cdots \times \left( \frac{1}{2}\,\mathcal{R}%
_{m_{2n-2}m_{2n-1}}^{n_{2n-2}n_{2n-1}}-t^{2}K_{m_{2n-2}}^{n_{2n-2}}K_{m_{2n-1}}^{n_{2n-1}}+%
\frac{t^{2}}{\ell ^{2}}\,\delta _{m_{2n-2}}^{n_{2n-2}}\delta
_{m_{2n-1}}^{n_{2n-1}}\right) \,.  \label{q_0}
\end{eqnarray}%
More explicitly, plugging in the components of the boundary curvature,%
\begin{equation}
\mathcal{R}_{m_{1}m_{2}}^{n_{1}n_{2}}=\frac{k}{r^{2}}\,\delta _{\lbrack
m_{1}m_{2}]}^{[n_{1}n_{2}]}\,,\qquad \mathcal{R}_{tm}^{tn}=0\,,  \label{calR}
\end{equation}%
the zero-point energy of the system is%
\begin{equation}
E_{vac}=2n\left( 2n-1\right) !c_{2n}\text{Vol}(\Gamma
_{2n-1})\,\lim_{r\rightarrow \infty }\,\int\limits_{0}^{1}dt\,t\,\left(
f^{2}-\frac{r\left( f^{2}\right) ^{\prime }}{2}\right) \left[ k+\left( \frac{%
r^{2}}{\ell _{e\!f\!f}^{2}}-f^{2}\right) \,t^{2}\right] ^{n-1}\,.
\label{Vac2n+1}
\end{equation}%
As the metric function and its derivative can be expanded as in Eqs.(\ref%
{f_asympt}) and (\ref{df_asympt}), we notice that all the terms that depend
on the parameter $\mu $ vanish in the limit $r\rightarrow \infty $. As
expected, the vacuum energy depends only on the topological parameter $k$,
the effective AdS radius and GB coupling, that is,%
\begin{eqnarray}
E_{vac} &=&\left( 2n-1\right) !\,c_{2n}\text{Vol}(\Gamma _{2n-1})\,k^{n}
\notag \\
&=&(-k)^{n}\frac{\text{Vol}(\Gamma _{2n-1})}{8\pi G}\,\ell
_{e\!f\!f}^{2n-2}\,\frac{(2n-1)!!^{2}}{(2n)!}\,\left( 1-\frac{2\alpha }{\ell
_{e\!f\!f}^{2}}\,\left( 2n-1\right) \left( 2n-2\right) \right) .
\end{eqnarray}%
The above formula matches the vacuum energy in EGB gravity obtained in Ref.%
\cite{Kofinas-Olea-GB} by means of Kounterterm regularization. This implies
that for an arbitrary NED Lagrangian the fall-off of the electromagnetic
field is always such that it does not contribute to the total energy of the
gravitational configuration.

\section{Conclusions}

We have used counterterms for Einstein-Gauss-Bonnet gravity coupled to
nonlinear electrodynamics in the form of polynomials in the extrinsic and
intrinsic curvatures of the boundary in order to regularize the conserved
charges in the AdS sector of the theory. It has been shown that this
regularization scheme (also known as Kounterterm method) provides finite
values for the mass for charged static black holes with spherical, locally
flat and hyperbolic transversal section in all dimensions, and the correct
vacuum energy in odd dimensions.

We have also analyzed the fall-off conditions that ensure the finiteness of
the electric charge for an arbitrary NED Lagrangian $\mathcal{L(}F^{2}%
\mathcal{)}$, which do not produce additional contributions to the mass of
black hole in Einstein-Gauss-Bonnet AdS gravity.

It is well-known that a vacuum energy for global AdS spacetime in odd
dimensions appears only in background-independent methods to compute
conserved quantities. This is particularly important from the semiclassical
point of view in order to interpret the Noether charges as thermodynamic
variables, and to consistently incorporate the vacuum energy in the
definition of internal energy of the system \cite{Miskovic-Olea Thermo
EGB-NED}, in a similar fashion as in Einstein-BI system \cite%
{Miskovic-OleaBI} (for a thermodynamic analysis of the same system
using a background-subtraction method see Ref.\cite{Banerjee}). The addition of a series of intrinsic counterterms in
pure EGB\ AdS gravity (see, eg., Refs.\cite{Brihaye-Radu,Astefanesei et
al,Liu-Sabra}) presents the advantage of obtaining the conserved quantities
from a boundary stress tensor, that is, as holographic charges. However, the
explicit form of such series does not exist for a high enough dimension. On
the contrary, an expression for the Kounterterms is given by Eqs.(\ref%
{B-even}) and (\ref{B_2n}) in all dimensions. In that respect, one would
like to see the above charges as coming from a quasilocal stress
(Brown-York) tensor. There are good reasons that make us think that this
could be possible, despite the fact that the on-shell variation of the
action takes the form
\begin{equation}
\delta I_{D}=\int\limits_{\partial \mathcal{M}}d^{D-1}x\,\sqrt{-h}\left(
\frac{1}{2}\,\tau _{i}^{j}\left( h^{-1}\delta h\right) _{j}^{i}+\Delta
_{i}^{j}\,\delta K_{j}^{i}+\Omega ^{i}\delta A_{i}\right) \,,
\label{arb varI}
\end{equation}%
where one cannot directly define a quasilocal stress tensor as $T^{ij}=(2/%
\sqrt{-h})\,(\delta I_{D}/\delta h_{ij})$.

Indeed, there are gravity theories where the surface term in $\delta I$
contains variations of the extrinsic curvature $\delta K_{j}^{i}$, which
cannot be eliminated by the addition of a generalized Gibbons-Hawking term,
and where a holographic stress tensor for AAdS spacetimes can be still read
off from the variation of the action. One example featuring this property is
Topologically Massive Gravity in 3D, where the surface term coming from the
variation of the gravitational Chern-Simons term contains $\delta K_{j}^{i}$%
. It is known that there is no term that can be added to the action to trade
it off by a piece along $\delta h_{ij}$. However, it can be shown that in
the asymptotically AdS sector of the theory, there is a contribution from
the gravitational Chern-Simons term to the holographic stress tensor which
couples to the conformal structure $g_{(0)ij}$, even though a quasilocal
stress tensor associated to $\delta h_{ij}$ cannot be defined \cite%
{Kraus-Larsen}. This follows from the fact that, for AAdS spaces, the
leading order in the expansion of the boundary metric is the same as the
leading order of the extrinsic curvature. A quasilocal stress tensor cannot
be identified either in 4D AdS gravity when one adds the (topological)
Gauss-Bonnet term to the Einstein-Hilbert action. In this case, the
Gauss-Bonnet term does not change the field equations in the bulk but, as
expected, it modifies the surface term in the variation of the action. In
this case, $\delta I$ also adopts the form of Eq.(\ref{arb varI}). However,
the second term in (\ref{gen GH}) --which in $D>4$ sets a well-defined
action principle when the metric is held fixed at the boundary -- cannot be
used for the same purpose in four dimensions. One can show that the
variation of the action produces a boundary stress tensor $\tau _{i}^{j}$
for AdS gravity (upon a suitable choice of the GB coupling) which is finite
and the same as the one prescribed by holographic renormalization \cite%
{Miskovic-Olea 4D}. This is a consequence of the fact that the contribution $%
\sqrt{-h}\,\Delta _{i}^{j}\,\delta K_{j}^{i}$ vanishes identically when one
performs an asymptotic expansion of the fields.

The above examples give some indication on what should be the pattern in
higher-dimensional Einstein-Hilbert and Einstein-Gauss-Bonnet AdS case: in $%
D=2n$ dimensions, the term that contains $\delta K_{j}^{i}$ should always
vanish as we approach to the asymptotic region, such that the quasilocal
stress tensor can be read off directly from Eq.(\ref{arb varI}). On the
other hand, in odd dimensions, $\Delta _{i}^{j}\,\delta K_{j}^{i}$ should
contribute with a finite piece to the holographic stress tensor which does
not modify the Weyl anomaly. We expect to provide a proof of the above claim
elsewhere.

\section*{Acknowledgments}

This work was funded by FONDECYT Grants 11070146, 1090357 and 1100755. O.M.
is supported by Project MECESUP UCV0602 and the PUCV through the projects
123.797/2007, 123.705/2010.

\appendix

\section{Kronecker delta of rank $p$ \ \label{Delta}}

The totally-antisymmetric Kronecker delta of rank $p$ is defined as the
determinant
\begin{equation}
\delta _{\left[ \mu _{1}\cdots \mu _{p}\right] }^{\left[ \nu _{1}\cdots \nu
_{p}\right] }:=\left\vert
\begin{array}{cccc}
\delta _{\mu _{1}}^{\nu _{1}} & \delta _{\mu _{1}}^{\nu _{2}} & \cdots &
\delta _{\mu _{1}}^{\nu _{p}} \\
\delta _{\mu _{2}}^{\nu _{1}} & \delta _{\mu _{2}}^{\nu _{2}} &  & \delta
_{\mu _{2}}^{\nu _{p}} \\
\vdots &  & \ddots &  \\
\delta _{\mu _{p}}^{\nu _{1}} & \delta _{\mu _{p}}^{\nu _{2}} & \cdots &
\delta _{\mu _{p}}^{\nu _{p}}%
\end{array}%
\right\vert \,.
\end{equation}
A contraction of $k\leq p$ indices in the Kronecker delta of rank $p$
produces a delta of rank $p-k$,
\begin{equation}
\delta _{\left[ \mu _{1}\cdots \mu _{k}\cdots \mu _{p}\right] }^{\left[ \nu
_{1}\cdots \nu _{k}\cdots \nu _{p}\right] }\,\delta _{\nu _{1}}^{\mu
_{1}}\cdots \delta _{\nu _{k}}^{\mu _{k}}=\frac{\left( N-p+k\right) !}{%
\left( N-p\right) !}\,\delta _{\left[ \mu _{k+1}\cdots \mu _{p}\right] }^{%
\left[ \nu _{k+1}\cdots \nu _{p}\right] }\,,
\end{equation}%
where $N$ is the range of indices.

\section{Hypergeometric function \label{Hyper}}

We use an integral representation of the Gauss' hypergeometric function,%
\begin{equation}
\left. _{2}F_{1}\right. (a,b;c;z)=\frac{\Gamma (c)}{\Gamma (b)\Gamma (c-b)}%
\int\limits_{0}^{1}du\,\frac{u^{b-1}\left( 1-u\right) ^{c-b-1}}{\left(
1-zu\right) ^{a}}\,,
\end{equation}%
where $c$ is not a negative integer and either $|z|<1$, or $|z|=1$ with $\Re
e(c-a-b)>0$. In particular, the following integral is solved in the text,
\begin{equation}
\int\limits_{0}^{1}du\,\frac{u^{b-1}}{\sqrt{1+zu}}=\frac{1}{b}\,\left.
_{2}F_{1}\right. \left( \frac{1}{2},b;b+1;-z\right) \,,\qquad b>0\,.
\label{integral}
\end{equation}%
The first derivative of the hypergeometric function is
\begin{equation}
\frac{d}{dz}\left. _{2}F_{1}\right. (a,b;c;z)=\frac{ab}{c}\,\left.
_{2}F_{1}\right. (a+1,b+1;c+1;z)\,,
\end{equation}%
and it expands for small $z$ as%
\begin{equation}
_{2}F_{1}(a,b;c;z)=1+\frac{ab}{c}\,z+\frac{a\left( a+1\right) b\left(
b+1\right) }{2c\left( c+1\right) }\,z^{2}+\mathcal{O}(z^{3})\,.
\end{equation}

\section{Gauss-normal coordinate frame \label{Gauss-normal}}

In Gaussian coordinates (\ref{radial foliation}), the only relevant
components of the connection $\Gamma _{\mu \nu }^{\alpha }$ are expressed in
terms of the extrinsic curvature $K_{ij}=-\frac{1}{2N}\,h_{ij}^{\prime }$ as
\begin{equation}
\Gamma _{ij}^{r}=\frac{1}{N}\,K_{ij\,},\qquad \Gamma
_{rj}^{i}=-NK_{j}^{i}\,,\qquad \Gamma _{rr}^{r}=\frac{N^{\prime }}{N}\,.
\label{KChr}
\end{equation}%
The radial foliation (\ref{radial foliation}) implies the Gauss-Codazzi
relations for the spacetime curvature, as well,
\begin{eqnarray}
R_{kl}^{ir} &=&\frac{1}{N}\,\left( \nabla _{l}K_{k}^{i}-\nabla
_{k}K_{l}^{i}\right) \,,  \label{Codazzi2} \\
R_{kr}^{ir} &=&\frac{1}{N}\,\left( K_{k}^{i}\right) ^{\prime
}-K_{l}^{i}\,K_{k}^{l}\,,  \label{Codazzi3} \\
R_{kl}^{ij} &=&\mathcal{R}_{kl}^{ij}(h)-K_{k}^{i}\,K_{l}^{j}+K_{l}^{i}%
\,K_{k}^{j}\,\equiv \mathcal{R}_{kl}^{ij}-K_{[k}^{[i}K_{l]}^{j]}\,,
\label{Codazzi1}
\end{eqnarray}%
where $\nabla _{i}=\nabla _{i}(h)$ is the covariant derivative defined in
the Christoffel symbol of the boundary $\Gamma _{ij}^{k}(g)=\Gamma
_{ij}^{k}(h)$ and $\mathcal{R}_{kl}^{ij}(h)$ is the intrinsic curvature of
the boundary.

\end{document}